%% file: Final-Manuscript.tex
\documentclass[11pt]{article}
 \usepackage{graphicx}

\usepackage{graphics,graphicx,psfrag,color,float, hyperref}
\usepackage{epsfig,bbold,bbm}
\usepackage{bm}
\usepackage{mathtools}
\usepackage{amsmath}
\usepackage{amssymb}
\usepackage{amsfonts}
\usepackage{amsthm, hyperref}
\usepackage{amsfonts}
\usepackage {times}
\usepackage{bbm}
\usepackage{soul}
\usepackage{titling}
\usepackage{amsmath}
\usepackage{layout}
\usepackage{amssymb}
\usepackage{amsfonts}
\usepackage{amsthm, hyperref}
\usepackage {times}
\usepackage{setspace}
\usepackage{amsfonts}
\newcommand{\Supp}{\operatorname{Supp}}

\newcommand{\beq}{\begin{equation}}
\newcommand{\enq}{\end{equation}}
\newcommand{\bel}{\begin{lemma}}
\newcommand{\enl}{\end{lemma}}
\newcommand{\bet}{\begin{theorem}}
\newcommand{\ent}{\end{theorem}}

\newcommand{\floor}[1]{\left\lfloor #1 \right\rfloor}

\newcommand{\eps}{\varepsilon}
\newcommand*{\cC}{\mathcal{C}}
\newcommand*{\cA}{\mathcal{A}}

\newcommand*{\cB}{\mathcal{B}}
\newcommand*{\cD}{\mathcal{D}}

\newcommand*{\cX}{\mathcal{X}}

\newcommand*{\cU}{\mathcal{U}}

\newcommand*{\bP}{\mathbf{P}}

\newcommand*{\Ib}{\bar{I}}

\newcommand*{\cY}{\mathcal{Y}}
\newcommand*{\hy}{\hat{y}}
\newcommand*{\hx}{\hat{x}}

\newcommand*{\bX}{\mathbf{X}}
\newcommand*{\bU}{\mathbf{U}}
\newcommand*{\bY}{\mathbf{Y}}
\newcommand*{\bQ}{\mathbf{Q}}

\newcommand*{\bx}{\mathbf{x}}

\mathchardef\mhyphen="2D

\newcommand*{\oH}{\overline{H}}
\newcommand*{\renyi}{R\'{e}nyi }

\makeatletter
\newcommand*{\rom}[1]{\expandafter\@slowromancap\romannumeral #1@}
\makeatother

\mathchardef\mhyphen="2D

\newtheorem{definition}{Definition}
\newtheorem{claim}{Claim}

\newtheorem{theorem}{Theorem}
\newtheorem{lemma}{Lemma}

\begin {document}
\title{Simple one-shot bounds for various source coding problems using smooth \renyi quantities}
\author{ 
Naqueeb Ahmad Warsi\\ 
%School of Technology and Computer Science\\
Tata Institute of Fundamental Research, Mumbai - 400 005, India\\
Email: warsi.naqueeb@gmail.com
}
\date{}
\maketitle

\begin{abstract}
We consider the problem of source compression under three different scenarios in the one-shot (non-asymptotic) regime. To be specific, we prove one-shot achievability and converse bounds on the coding rates for distributed source coding, source coding with coded side information available at the decoder and source coding under maximum distortion criterion. The one-shot bounds obtained are in terms of smooth max \renyi entropy and smooth max \renyi divergence. Our results are powerful enough to yield the results that are known for these problems in the asymptotic regime both in the i.i.d. (independent and identically distributed) and non-i.i.d. settings.
\end{abstract}
\section{Introduction}
In this manuscript  we study three source coding problems in the one-shot regime.

\paragraph{Distributed source coding (the Slepian-Wolf problem):} In this problem, two different correlated random variables are encoded separately and the decoder is able to simultaneously decode the correlated random variables with high probability. Formally, this problem is defined as follows. There are three parties: Alice, Bob and Charlie. Alice possesses the random variable $X$ and Bob possesses the random variable $Y$. Alice and Bob both want to communicate their respective random variable to Charlie. We assume here that Alice and Bob are separated in space and do not collaborate while communicating their share of the pair $(X,Y)$. To accomplish this task Alice sends a message $M_A$ of $\ell_A$ bits to Charlie and Bob sends a message $M_B$ of $\ell_B$ bits to Charlie. On receiving the message pair $(M_A, M_B)$ Charlie tries to  simultaneously reconstruct the pair $(X,Y)$. We are required to determine the conditions on $\ell_A$ and $\ell_B$ so that Charlie may reconstruct $(X,Y)$ simultaneously using $(M_A, M_B)$ with probability at least $1-\varepsilon$. Figure \ref{distributed} below illustrates the problem of distributed source coding. 
\begin{figure}[H]
\centering
\vspace{0.8cm}
\resizebox{0.45\textwidth}{!}{
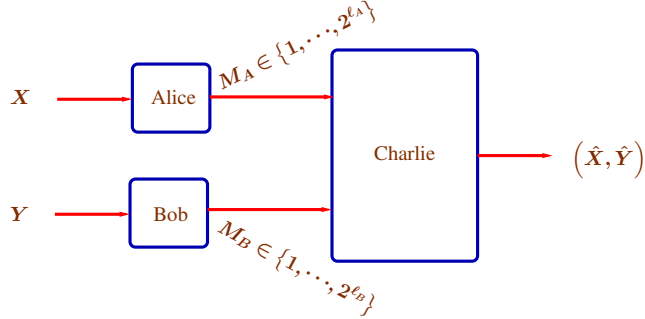
}
\vspace{0.6cm}
\caption{Distributed source coding.}
\label{distributed}
\end{figure}
We prove the achievability and converse bounds for this problem. A converse bound for this problem also appears in \cite{sharma-warsi-allerton}. However, we give a different proof. Our bounds are stated in terms of smooth zeroth order \renyi entropy (see Definition \ref{smcond} for the definition). We prove the following results.
\begin{theorem}
\label{thasw}
(Achievability) Let $(X,Y)$ be a pair of random variables taking values in
$\mathcal{X} \times \mathcal {Y}$ with some joint distribution $p_{XY}$ and let $\eps \in (0,1)$ be given. Furthermore, let
$\ell_A, \ell_B$ be such that 
\begin{align*}
\ell_A & \geq H_{0}^{\frac{\eps}{6}}[X|Y]-\log\left(\frac{\eps}{ 6}\right)\\
\ell_B& \geq H_{0}^{\frac{\eps}{6}}[Y|X]-\log\left(\frac{\eps}{ 6}\right)\\
\ell_A+ \ell_B & \geq H_{0}^{\frac{\eps}{6}}[XY]-\log\left(\frac{\eps}{ 6}\right).
\end{align*}
Then, there exist encoding functions $e_A: \mathcal{X} \rightarrow
[1:2^{\ell_A}]$ and $e_B: \mathcal{Y} \rightarrow [1:2^{\ell_B}]$ and a
decoding function $d_C: [1:2^{\ell_A}] \times [1:2^{\ell_B}] \rightarrow
\mathcal{X} \times \cY$ such that $\Pr\left\{(X,Y) \neq d_C(e_A(X), e_B(Y))\right\} \leq \eps.$
\end{theorem} 
\begin{theorem}
\label{thcsw}
(Converse) Suppose $(X,Y)$ be a pair of random variables taking values in $\cX \times \cY$, such that for some encoding functions $e_A: \mathcal{X} \rightarrow [1:2^{\ell_A}]$, $e_B: \mathcal{Y} \rightarrow [1:2^{\ell_B}]$ and $d_{C}:[1:2^{\ell_A}] \times [1:2^{\ell_B}] \rightarrow \cX\times \cY$, we have $\Pr\left\{(X,Y) \neq d_{C}(e_A(X), e_B(Y))\right\} \leq \eps$, where $\eps \in (0,1)$. Then,
\begin{align*}
\ell_A & \geq H_{0}^{\eps}[X|Y]\\
\ell_B & \geq H_{0}^{\eps}[Y|X]\\
\ell_A + \ell_B  & \geq H_{0}^{\eps}[XY].
\end{align*}
\end{theorem}
{\bf{Related work:}}
Similar results were also independently derived by Uyematsu and Mastsuta in \cite[Theorem 2]{uteymatsu-matsuta-2014}. However, the bounds obtained by us is somewhat better than the bounds obtained in \cite{uteymatsu-matsuta-2014}. This is because $H^{\eps}_0$ is somewhat smaller than the corresponding quantity used in \cite{uteymatsu-matsuta-2014}, however, the difference is very minor. The proof techniques for deriving the achievability part remains the same both in our work and in \cite{uteymatsu-matsuta-2014}. However, the proof techniques used by us for deriving the converse part is different from that used in \cite{uteymatsu-matsuta-2014}. 

In the asymptotic iid setting, the distributed source coding problem was studied by Slepian and Wolf \cite{slepian-wolf-1973}, who showed the following:
\begin{theorem}
\label{slepian-wolf}
The optimal rate region for distributed source coding of a two discrete memoryless source $\left(\cX\times\cY, p_{XY}\right)$ is the set of rate pairs $(R_A, R_B)$ such that
\begin{align*}
R_A&\geq H[X|Y]\\
R_B&\geq H[Y|X]\\
R_A+R_B&\geq H[XY].
\end{align*}
\end{theorem}
\paragraph{Source coding with coded side information available at the decoder:} The setting is similar to one above. The only difference is that Charlie must now try and recover $X$ alone. Figure \ref{helper11} below illustrates this problem.  
\begin{figure}[H]
\centering
\vspace{0.7cm}
\resizebox{0.45\textwidth}{!}{
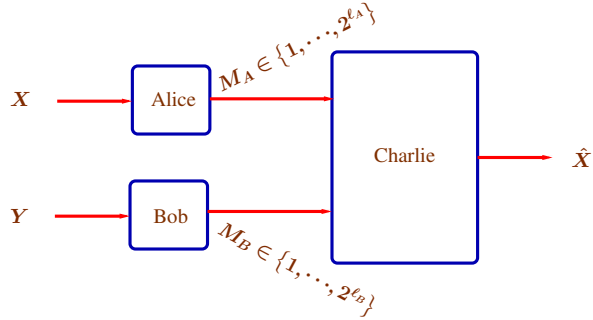
}
\vspace{0.6cm}
\caption{Source coding with a helper}
\label{helper11}
\end{figure}
We give two different achievability bounds along with the converse bounds for this problem. The proof for each of these bounds are based on completely different techniques. Our bounds are in terms of smooth zeroth order \renyi entropy and smooth max \renyi divergence (see Defintion \ref{smcond} and Definition \ref{smoothorderinf} for the respective definitions). We prove the following:
\begin{theorem}\label{oneshotsideinf}(Achievability)
Let $(X,Y)$ be a pair of random variables taking values in
$\mathcal{X} \times \mathcal {Y}$ with some joint distribution and let $\eps \in (0,1)$ be given. Furthermore, let
$\ell_A, \ell_B, \varepsilon_A, \varepsilon_B$ be such that for some
random variable $U$ satisfying $X$---$Y$---$U$, we have
\begin{align*}
\ell_{A} & \geq H^{\eps_{A}}_{0}[X|U] - \log (\eps -\eps_B)\\
\label{swr2}
\ell_{B} & \geq I^{\bar{\eps}_{B}}_{\infty}[U;Y]+\log[-\ln (\eps_B-\bar{\eps}_{B}-2\eps_{A}^{\frac{1}{2}}) ],
\end{align*}
where $\eps_A+\eps_B \leq \eps;$ $\bar{\eps}_{B}+ 2\eps_{A}^{\frac{1}{2}} < \eps_B$ and $I^{\bar{\eps}_{B}}_{\infty}[U;Y] \geq 0.$ Then, there exist encoding functions $e_A: \mathcal{X} \rightarrow
[1:2^{\ell_A}]$ and $e_B: \mathcal{Y} \rightarrow [1:2^{\ell_B}]$ and a
decoding function $d_C: [1:2^{\ell_A}] \times [1:2^{\ell_B}] \rightarrow
\mathcal{X}$ such that 
$\Pr\left\{X \neq d_C(e_A(X), e_B(Y))\right\} \leq \eps.$
\end{theorem}
\begin{theorem}\label{jaikumar's proof}
(Alternate achievability bounds)
Let $(X,Y)$ be a pair of random variables taking values in
$\mathcal{X} \times \mathcal {Y}$ with some joint distribution and let $\eps \in (0,1)$ be given. Furthermore, let
$\ell_A, \ell_B, \varepsilon_A, \varepsilon_B$ be such that for some
random variable $U$ satisfying $X$---$Y$---$U$, we have
\begin{eqnarray}
\ell_A &\geq& H_0^{\varepsilon_A}[X|U] + \log(1/\varepsilon_A);\\
\label{altb1}
\ell_B &\geq& \max\left\{0, I_{\infty}^{\varepsilon_B}[U;Y] +1 \right\}+ \log \ln (1/\varepsilon_B),
\label{altb2}
\end{eqnarray}
where $2
\varepsilon_A + 4 \varepsilon_B \leq \eps$. Then, there exist encoding functions $e_A: \mathcal{X} \rightarrow
[1:2^{\ell_A}]$ and $e_B: \mathcal{Y} \rightarrow [1:2^{\ell_B}]$ and a
decoding function $d_C: [1:2^{\ell_A}] \times [1:2^{\ell_B}] \rightarrow
\mathcal{X}$ such that 
\begin{equation}
\Pr\left\{X \neq d_C(e_A(X), e_B(Y))\right\} \leq \eps. \label{eq:prob}
\end{equation}
\end{theorem}

On the way of proving Theorem $5$ we prove the following lemma which we believe could be of independent interest.

{\bf{Lemma. }}{\emph{Let $U \sim p_U$ take values over the set ${\mathcal U}$.  Let $L$
be a positive integer. Let $R$ be distributed uniformly in the real unit
cube $[0,1]^{L+1}$.  Then, there is a decoding function $d: [0,1]^{L+1}
\times [1:L] \rightarrow {\mathcal U}$ such that for every distribution
$q$ on ${\mathcal U}$, there is an encoder $e_q: [0,1]^{L+1}
\rightarrow [1:L]$, such that the random variable $V= d(R,e_q(R))$ (notice the dependence of $V$ on $q$)
satisfies $\|p_{V} - q\| \leq 2\exp\left(- L2^{-D_{\infty}(q \|
p_U)}\right)$}}.
\begin{theorem}
\label{conversesideinf}
(Converse) Suppose $(X,Y)$ be a pair of random variables taking values in $\cX \times \cY$, such that for some encoding functions $e_A: \mathcal{X} \rightarrow [1:2^{\ell_A}]$, $e_B: \mathcal{Y} \rightarrow [1:2^{\ell_B}]$ and $d_{C}:[1:2^{\ell_A}] \times [1:2^{\ell_B}] \rightarrow \cX$, we have $\Pr\left\{X \neq d_{C}(e_A(X), e_B(Y))\right\} \leq \eps$, where $\eps \in (0,1)$. Then,
\begin{align*}
\ell_A &\geq H^{\eps}_{0}[X|U]\\
\ell_B  & \geq I^{\eps}_{\infty}[U;Y]+ \log (\eps),
\end{align*}
where 
$U =e_B(Y).$
\end{theorem}

In the asymptotic iid setting this problem was studied by Wyner \cite{wyner-1975}, who showed the following:
\begin{theorem}
\label{helper}
Let $(X,Y) \sim p_{XY}$ be a two discrete memoryless correlated sources. The optimal rate region for lossless source coding with a helper is the set of rate pairs $(R_A, R_B)$ such that
\begin{align*}
R_A &\geq H[X|U]\\
R_B & \geq I[U;Y],
\end{align*}
for some $p_{U|Y}$.
\end{theorem}
\paragraph{Source coding under maximum distortion criterion:}
In this problem there are two parties Alice and Bob. Alice possesses a random variable $X$. Alice wishes to send a message $M_A$ of $\ell_A$ bits to Bob so that Bob can construct $Y$ from the message $M_A$ such that  $\bar{\lambda}_{\eps}(X,Y) \leq \gamma$, where 
\beq
\label{d}
\bar{\lambda}_{\eps}(X,Y) := \inf\bigg\{\lambda: \Pr\{d(X,Y) \leq \lambda\} > 1- \eps\bigg\}, 
\enq
where $d: \cX\times\cY \to \mathbb{R^+}$ and we assume that $d$ is a bounded function. Figure \ref{121} below illustrates this problem.
\begin{figure}[H]
\centering
\vspace{0.7cm}
\resizebox{0.55\textwidth}{!}{
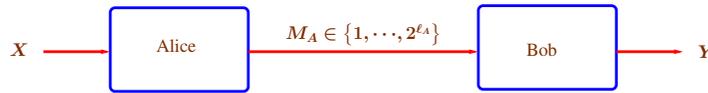
}
\vspace{0.6cm}
\caption{Source coding under distortion criterion}
\label{121}
\end{figure}
We prove the following achievability and converse bounds for this problem. Our results are stated in terms of smooth max \renyi divergence (see Defintion\ref{smoothorderinf} for the definition of smooth max \renyi divergence).
\begin{theorem}\label{distac}(Achievability)
Let $X \sim p_{X}$, $\eps \in (0,1)$ and a bounded distortion function $d: \cX \times \cY \to \mathbb{R}^+$  be given. Furthermore, let $\ell_A$ be such that 
\beq
\ell_A \geq \max\left\{0,I^{\eps_1}_{\infty}[X;Y]\right\} +\log[-\ln(\eps-2\eps_1)], \nonumber
\enq
where $2\eps_1 < \eps$ and the random variable $Y$ is such that $\bar{\lambda}_{\eps_1}(X,Y) \leq \gamma$ (see \eqref{d} for the definition). Then there exists encoding function $e:\cX \to [1:2^{\ell_A}]$ and a decoding function $d: [1:2^{\ell_A}] \to \cY$ such that $\bar{\lambda}_{\eps}(X,d(e(X))) \leq \gamma.$
\end{theorem}
\begin{theorem} (Converse)
\label{distconv}
Let $X\sim P_X$ with range $\cX$. If any encoding function $e: \cX \to [1:2^{\ell_A}]$ and decoding function $f : [1:2^{\ell_A}] \to \cY$ satisifies
\beq
\bar{\lambda}_{\eps}(X,f(e(X))) \leq \gamma. \nonumber
\enq
Then,
$
\ell_A \geq  I^{\eps}_{\infty}[X;Y] + \log \eps, 
$
where $\eps \in (0,1).$
\end{theorem}

The optimal rate for this problem in the asymptotic setting was proved by Steinberg and V\'{e}rdu \cite[Theorem 5.4.1]{han}. To state their result we need the following definitions. 
\begin{definition}
Let $(\bX,\bY)$ be a sequence of pair of random variables, where for every $n$ $(X^n,Y^n) \sim p_{X^nY^n}$ and take values over the set $(\cX^n \times \cY^n)$. Let $\left\{d_n\right\}_{n=1}^{\infty}$ be a sequence of bounded functions where for every $n$, $d_n: \cX^n \times \cY^n \to \mathbb{R}^+$. The maximum distortion $\bar{\lambda}(\bX, \bY)$ between the random $\bX$ and $\bY$ is defined as follows:
\beq
\bar{\lambda}(\bX, \bY):= \inf\left\{\lambda: \lim_{n \to \infty }\Pr \left\{d_n(X^n,Y^n) > \lambda \right\} = 0\right\}.
\enq
\end{definition}
\begin{definition}
Let $(\bX,\bY):=\left\{X^n,Y^n\right\}_{n=1}^{\infty}$ be a sequence of pair of random variables, where for every $n$ $(X^n,Y^n) \sim p_{X^nY^n}$ and take values over the set $(\cX^n \times \cY^n)$. The spectral sup mutual information rate $\overline{I}(\bX;\bY)$ between $\bX$ and $\bY$ is defined as follows:
\beq
\overline{I}(\bX;\bY):= \inf\left\{a: \lim_{n \to \infty }\Pr \left\{\frac{1}{n}\log\frac{p_{X^nY^n}}{p_{X^n} p_{Y^n}} > a\right\} = 0\right\},
\enq
where the probability above is calculated with respect to $p_{X^nY^n}$.
\end{definition}
Using the above definitions we now state the result of Steinberg and V\'{e}rdu \cite[Theorem 5.4.1]{han}. 
\begin{theorem}
Let $\bX$ be a sequence of arbitrarily distributed random variables. For any distortion measure $d_n$ the optimal rate $R_A$ satisfies
\beq
\label{spectd}
R_A= \inf_{\bY: \bar{\lambda}(\bX, \bY) \leq \lambda}\overline{I}(\bX;\bY),
\enq
where the infimum on the right-hand side is taken with respect to the all sequences ${\bY}$ satisfying
$\bar{\lambda}(\bX, \bY) \leq \lambda$.
\end{theorem}
\section{One-shot paradigm}
All our contributions are in the one-shot paradigm, and it will be useful to place our results in context by stating a prototype result in the literature. Renner and Wolf \cite{renner-wolf-2004} show one-shot bounds on the coding rate for source compression. The problem of one-shot source compression is the following. There are two parties Alice and Charlie. Alice possesses a random variable $X$ and wishes to communicate $X$ to Charlie. To accomplish this task Alice sends a message $M_A$ of $\ell_A$ bits to Charlie. On receiving the message $M_A$ Charlie tries to reconstruct the random variable $X$. We are required to determine the conditions on $\ell_A$ so that Charlie may reconstruct $X$ using $M_A$ with probability at least $1-\eps$. Renner and Wolf prove the following achievability and converse results for this problem \cite[Theorem 7.8]{renner-wolf-2004}.
\begin{theorem}
\begin{description}
\item[$(\mbox{Achievability})$] Let $X \sim p_X$ be a random variable taking values over the set $\cX$ and let $\eps \in (0,1)$ be given.  Furthermore, let
\begin{align*}
\ell_A \geq H^{\frac{\eps}{2}}_{0}[X] - \log \eps/2.
\end{align*}
Then, there exist an encoding function $e_A : \cX \to [1:2^{\ell_A}]$ and decoding function $d_C : [1:2^{\ell_A}] \to \cX$ such that $\Pr \left\{X \neq d_C\left(e_A(X)\right)\right\} \leq \eps$.
\item[$(\mbox{Converse})$] Let $X \sim p_X$ be a random variable taking values over the set $\cX$. If any encoding function $e_A : \cX \to [1:2^{\ell_A}]$ and decoding function $d_C : [1:2^{\ell_A}] \to \cX$ satisfies $\Pr \left\{X \neq d_C\left(e_A(X)\right)\right\} \leq \eps$. Then $\ell_A \geq H^\eps_{0}[X]$.
\end{description}
\end{theorem}
In traditional information theory literature it is common to study the
underlying problems in the asymptotic setting, often assuming that the
channel characteristics do not change over multiple use. The proofs
appeal to {\em typicality} of sequences: the empirical
distribution of symbols in a long sequence of trials will with high
probability be close to the true
distribution~\cite{covertom}.  However, information
theoretic arguments based on typicality or the related Asymptotic
Equipartition Property (AEP) assume that both the source and channel
are stationary and/or ergodic (memoryless), assumptions that are not
always valid, for example, in \cite{gray-book} Gray analyzes the details of asymptotically mean stationary sources, which are neither stationary nor ergodic. 

To overcome the above mentioned assumptions there have been considerable interest in one-shot information theory in the recent past. In one-shot we only have one instance of the source and/or allowed to use the channel only once  
further we are allowed only a small probability of error. Bounds proved in
such scenarios are said to hold in the one-shot setting. Such results
are more general, for one can always view a channel used repeatedly as
a single channel with a larger alphabet, and recover the asymptotic
bounds as a special case. The general one-shot results, while
sometimes technically harder to show (for arguments based on
typicality is no longer available), are often so strong that
nothing is lost in deriving the asymptotic results from them. That is,
the proofs then have the following structure: (1) Show one-shot bounds
to derive bounds in terms of one-shot versions of information
theoretic quantities; (2) Depending on the asymptotic setting under
consideration, invoke existing convergence results in the literature
and conclude that the same bounds must hold there. In \cite{renes-renner-2011} Rennes and Renner proved one-shot bounds on the private capacity of wiretap channel using smooth \renyi entropies. We quote Rennes and Renner from \cite{renes-renner-2011} wherein they mention that, ``One-shot scenario of structureless resources, meaning the coding scheme does not rely on repeated uses of a memoryless channel. Rather, the one-shot scenario is considerably broader in approach, encompassing not only channels in the traditional sense of communication (both with and without memory), but also channels as models for the dynamics of a physical system, for which memoryless  assumption would be out of place". There have been several such one-shot results in the literature. See for example \cite{renner-isit-2009, verdu-2012, Polyanskiy-Poor-Verdu-2012} and references therein.

Apart from the classical case the quantum equivalent of smooth \renyi quantities have been extensively used to prove one-shot bounds for various quantum information theoretic protocols. See for example  \cite{datta-renner-2009, konig-op-mean-2009,dupuis-broadcast-2010,berta-reverse-shannon-2011, 
datta-fqsw-2011,renes-renner-2011, wang-renner-prl, J.M.Renes, datta-renner-wilde-2013} and references therein. 

In this manuscript we carry forward the idea of using smooth \renyi quantities to prove one-shot achievability and converse bounds for three different source coding problems namely, the Slepain-Wolf problem, source coding with helper and source coding under maximum distortion criterion. For the distributed source coding problem the techniques involved for proving one-shot bounds are simple and they only involve exploiting the definition of smooth zeroth order \renyi entropy. However, for the source coding with the helper problem the techniques involved are not simple and requires new techniques and considerable mathematical manipulations. Further, we give two different achievability bounds for this problem. The techniques involved for proving each of these bounds are considerably different from each other.  

\section{Notation and Definition}
\label{notation and definitions}
We will assume that all the random variables are discrete and have finite range. We represent a random sequence of length $n$ by $X^n$ and a particular realization of $X^n$ by $x^n$. Notation $\bX$ will be used to represent an arbitrary sequence of random variables, i.e., $\bX = \left(X_1, X_2, \cdots\right)$. We use the notation $|\cdot|$ to represent the cardinality of a set. The set $\{x^n : p_{X^n}(x^n)>0\}$ is denoted by $\mbox{Supp} (p_{X^n})$. The notation $X$---$Y$---$Z $ will be used to represent the fact that the random variables $X$, $Y$ and $Z$ form a Markov chain. $\cX \times \cY$ will represent the cartesian product of two sets. Similarly, $(\cX \times \cY)^n$ will represent the $n\mhyphen$th Cartesian product of the set $\cX \times \cY$. The notation $\mathbb{N}$ is used to represent the set of natural numbers. The $L_1$ distance between any two probability mass functions defined on a set $\cX$ will be represented as
\beq
\|P-Q\|:= \sum_{x\in\cX}|P(x)-Q(x)|. \nonumber
\enq
We will assume throughout our discussions in this paper that $\log$ is to the base 2. The notation $[1:L]$ will be used to represent the set $\left\{1,\cdots, L\right\}.$
\begin{definition}({{Zeroth order \renyi entropy}} \cite{renyi-1960})\\
\label{zorderrenyi}
Let $X \sim p_{X}$, with range $\cX$. The zero order \renyi entropy \cite{renyi-1960} of $X$ is defined as
\beq
H_{0}(X) = \log |\Supp(p_X)|. \nonumber
\enq
\end{definition}
\begin{definition}({Zeroth order conditional \renyi entropy}  \cite{renner-wolf-2004})\\
\label{condzero}
Let $(X,Y) \sim p_{XY}$, with range $\cX \times \cY$. The zero order conditional smooth \renyi entropy  of $X$ given $Y$ is defined as
\begin{align*}
H_{0}(X|Y) := \log\max_{y\in\cY} |\Supp(p_{X|Y =y})|,
\end{align*}

\end{definition}

\begin{definition} \label{smoothdef1} ({Zeroth order smooth \renyi entropy \cite{renner-wolf-2004}})\\
Let $X \sim p_X$, with range $\cX$. For $\eps \in (0,1)$, the zero order smooth \renyi entropy of a random variable $X$ is defined as 
\beq
\label{smin}
H_{0}^{\eps} (X) :=  \inf_{q \in \cB^{\eps}(p_X)} \log |\Supp(q)|,
\enq
where
$ \cB^{\eps}(p_X) = \bigg\{q : 0\leq q(x) \leq p_X(x),\forall x \in \cX~\mbox{and}~\sum_{x\in \cX} q(x) \geq 1-\eps\bigg\}.
$
\end{definition}
\begin{definition}({Zeroth order conditional smooth \renyi entropy \cite{renner-wolf-2004}})\\
\label{smcond}
Let $(X,Y) \sim p_{XY}$, with range $\cX \times \cY$. For $\eps \geq 0$, the zero order conditional smooth \renyi entropy  of $X$ given $Y$ is defined as
\begin{align}
\label{condsmmax}
H_{0}^{\eps} (X|Y) := \inf_{q \in \cB^{\eps}(p_{XY})} \log\max_{y\in\cY} |\Supp(q(X|Y =y))|,
\end{align}
where 
$
q{(X=x|Y=y)}  =
\begin{cases}
\frac{q(x,y)}{p_Y(y)} & \mbox{if }~ p_Y(y) > 0 ; \\
0        & \mbox{otherwise.}
\end{cases}
$ and $ \cB^{\eps}(p_{XY}) = \bigg\{q : 0\leq q(x,y) \leq p_{XY}(x,y),\forall (x,y) \in \cX \times\cY~\mbox{and}~\sum_{(x,y)\in \cX\times\cY} q(x,y) \geq 1-\eps\bigg\} 
.$ 
\end{definition}
\begin{definition} ({Limit inferior in probability} \cite{han})\\
\label{definf}
Let $\{X_n\}_{n=1}^{\infty}$ be an arbitrary sequence of random variables. Then 
\beq
p \mhyphen\liminf_{n\to\infty}X_n:= \sup\left\{\alpha\big|\limsup_{n\to\infty}\Pr\{X_n <\alpha\} = 0 \right\}.
\enq
\end{definition}

\begin{definition}({Limit superior in probability} \cite{han})\\
\label{defsup}
Let $\left(X_1, X_2, \cdots,\right)$ be an arbitrary sequence of random variables. Then 
\beq
p \mhyphen\limsup_{n\to\infty}X_n:= \inf\left\{\alpha\big|\liminf_{n\to\infty}\Pr\{X_n < \alpha\} = 1 \right\}.
\enq
\end{definition}

\begin{definition}(Spectral sup$\mhyphen$entropy rate \cite{han})\\
The spectral sup$\mhyphen$entropy rate of $\bX$ is denoted by $\oH[\bX]$ and is defined as
\beq
\oH[\bX]:= p \mhyphen\limsup_{n\to\infty}\frac{1}{n} \log \frac{1}{p_{X^n}(X^n)}.
\enq
\end{definition}

\begin{definition}(Smooth max \renyi divergence)\\
\label{smoothorderinf} 
Let $P$ and $Q$ be two probability mass functions on the set $\cX$ such that $\Supp(P) \subseteq \Supp(Q)$.  The smooth max \renyi divergence between $P$ and $Q$ for $\eps \in [0,1)$ is defined as
\beq
\label{sminf}
D^{\eps}_{\infty}(P||Q) := \log \inf_{\phi \in \cB^{\eps}(P)} \max_{x : P(x)>0} \frac{\phi(x)}{Q(x)}, 
\enq
where
$ \cB^{\eps}(P) = \bigg\{\phi : 0\leq \phi(x) \leq P(x),\forall x \in \cX~\mbox{and}~\sum_{x\in \cX} \phi(x) \geq 1-\eps\bigg\}.
$ Notice that $\cB^{\eps}(P)$ also contains functions which are not probability mass functions, therefore smooth max \renyi divergence can be negative. Also, $D^{\eps}_{\infty}(P||Q)$ is a non-increasing function of $\eps$ and for $\eps=0$ it reduces to max \renyi divergence.

Let $(U,Y) \sim p_{UY}$. We will use the notation $I^{\eps}_{\infty}[U;Y]$ to represent $D^\eps_{\infty}(p_{UY} \| p_U \times p_Y).$
\end{definition}

\begin{definition}({Specrtal sup-mutual information rate} \cite{han})\\
\label{asyminftydiv}
Let $\bP = \{P_n\}_{n=1}^{\infty}$ and $\bQ = \{Q_n\}_{n=1}^{\infty}$ be an arbitrary sequences of probability mass functions defined on the set $\{\cX^n\}_{n=1}^{\infty}$, where $\cX^n$ is the $n\mhyphen$th cartesian product of the set $\cX$ and $|\cX| < \infty$. Assume that for every $n \in \mathbb{N}$, $\Supp(P_n) \subseteq \Supp(Q_n)$. The spectral sup-mutual information rate between $\bP$ and $\bQ$ is defined as follows
\beq
\label{specsup}
\Ib[\bP;\bQ] := p \mhyphen\limsup_{n\to\infty}\frac{1}{n} \log \frac{P_n}{Q_n}, \nonumber
\enq 
where the probability on the R.H.S. of the above equation is calculated with respect to the measure $P_n$.
\end{definition}

\section{Distributed Source Encoding of Correlated sources}

\subsection{Proof of Theorem \ref{thasw}}
Let $\ell_A$ and $\ell_B$ satisfy the assumptions of the Theorem. We now show the existence of suitable encoding and decoding functions.

{\bf{Random code generation:}} Randomly and independently assign  an index $i \in [1: 2^{\ell_A}]$ to every realization $x \in \cX$. Let $\Gamma_{X} (i)$ be the set of the realizations of the random variable  $X$ which have been assigned the index $i$. Similarly, to every realization $y \in \cY$ we assign an index $j \in [1: 2^{\ell_B}]$. Let $\Gamma_{Y} (j)$ be the set of the realizations of the random variable  $Y$ which have been assigned the index $j$. Reveal the bin assignments to the encoders and the decoder.

{\bf{Encoding:}} If Alice observes a realization $x \in \Gamma_X (i)$, then she transmits $i$ (this defines the encoding function $e_A$). In a similar way if Bob observes a realization $y \in  \Gamma_{Y}(j)$, then he transmits $j$ (this defines the function $e_B$).

{\bf{Decoding:}} Before giving charlie's strategy let us first define $q \in \cB^{\frac{\eps}{2}}(p_{XY})$ such that 
\begin{align}
\label{bsw1}
H_{0}^{\frac{\eps}{6}}[X Y] &  \geq \log|\mbox{Supp} (q(X, Y))|\\
\label{bsw2}
H_{0}^{\frac{\eps}{6}}[X|Y]& \geq \max _{y \in \cY}\log|\mbox{Supp} (q(X|Y=y))|\\
\label{bsw3}
H_{0}^{\frac{\eps}{6}}[Y|X] & \geq \max _{x \in \cX}\log|\mbox{Supp}(q(Y|X=x))|.
\end{align}
The existence of such a $q$ follows from Lemma \ref{swa} given at the end of the subsection. Let $(i,j)$ be the index pair received by the decoder. Charlie declares $(\hx, \hy)$ to be the estimate of the source pair if this pair is a unique pair in the bin $\Gamma_{X}(i) \times \Gamma_{Y}(j)$ such that $q(\hx,\hy)>0$. Otherwise the decoder declares an error (this defines the decoding function $d_C$). 

{\bf{Probability of error:}} We give an upper bound on the probability of error averaged over bin assignments. This would imply that the same upper bound will  hold for some fixed bin assignments for random $X$ and $Y$. Let $I$ and $J$ denote the random indices for $X$ and $Y$. Charlie makes an error if and only if one of the following events occur:
\begin{align*}
E_1 &:= \big\{(X,Y) \notin \mbox{Supp}(q)\big\}\\
E_2 &:= \big\{\exists x^\prime \neq X : x^\prime \in \Gamma_X(I) ~ \mbox{and} ~q(x^\prime,Y)>0\big\}\\
E_3 &:= \big\{\exists y^\prime \neq Y : y^\prime \in \Gamma_{Y}(J) ~ \mbox{and} ~q(X, y^\prime)>0\big\}\\
E_4 &:= \big\{\exists x^\prime \neq X, \exists y^\prime \neq Y : x^\prime \in \Gamma_{X}(I), y^\prime \in \Gamma_{Y}(J)~ \mbox{and} ~q(x^\prime, y^\prime)>0\big\}.
\end{align*}
Thus,
\begin{align*}
P_{e} &= \Pr\{E_1 \cup E_2\cup E_3\cup E_4\}\\
& \leq \Pr\{E_1\}+\Pr\{E_2\}+\Pr\{E_3\}+\Pr\{E_4\},
\end{align*} 
where the above inequality follows from the union bound. We now bound $\Pr\{E_1\}$, as follows
\begin{align*}
\Pr\{E_1\} &= \sum_{(x,y) \in \cX \times \cY: q(X,Y) = 0}p_{XY}(x,y)\\
&\overset{a} \leq \frac{\eps}{2} ,
\end{align*} 
where $a$ follows from the fact that for every $(x,y) \in \mbox{Supp}(q)$, $q(X,Y) = p_{XY}(x,y)$ and $q \in \cB^{\frac{\eps}{2}}(p_{XY})$. We now bound $\Pr\{E_2\}$ as follows.
\begin{align*}
\Pr\{E_2\}
& = \sum_{(x,y) \in \cX \times \cY} \Pr\big\{(X,Y) = (x,y)|X \in \Gamma_X(1)\big\} \Pr\big\{\exists x^\prime \neq  x: x^\prime \in \Gamma_X(1)  ~\mbox{and} \\
&\hspace{27mm}q(x^\prime,y)>0\big| x \in\Gamma_{X}(1), (X,Y) = (x,y)\big\}\\
& \leq \sum_{(x,y) \in \cX \times \cY}p_{XY}(x,y) \sum_{\substack {x^\prime \neq x \\ q(x^\prime,y) > 0}}\Pr (x^\prime \in \Gamma_X(1))\\
& \overset{a} \leq 2^{-\ell_A}\sum_{(x,y) \in \cX \times \cY}p_{XY}(x,y)\max_{y \in \cY} |\mbox{Supp}(q(X|Y=y))| \\
& = 2^{-\ell_A}\max_{y \in \cY} |\mbox{Supp}(q(X|Y=y))| \\
& \overset{b}\leq 2^{-\ell_A+H_{0}^{\frac{\eps}{6}}[X|Y]},
\end{align*}
where $a$ follows because of the randomness in the code and the fact that we have randomly and independently assigned an index $i \in [1: 2^{\ell_A}]$ to every realization $x \in \cX$ and $b$ follows from \eqref{bsw2}. Using similar arguments we have the following bounds
\beq
\Pr\{E_3\} \leq 2^{-\ell_B + H_{0}^{\frac{\eps}{6}}[Y|X]}, \nonumber
\enq 
and
\beq
\Pr\{E_4\} \leq 2^{-\left(\ell_A+ \ell_B\right)+H_{0}^{\frac{\eps}{6}}[XY]}. \nonumber
\enq
It now follows from our assumptions about $\ell_A$, $\ell_B$ and union bound that $P_e \leq \eps$.
This completes the proof.\\
\begin{lemma}
\label{swa}
Let $(X,Y) \sim p_{XY}$ with range $\cX \times \cY$ and let $\eps \in (0,1)$. There exists a positive function $q \in \cB^{\frac{\eps}{2}}(p_{XY})$ such that
\begin{align*}
H_{0}^{\frac{\eps}{6}}[XY] &  \geq \log|\Supp (q(X, Y))|,\\
H_{0}^{\frac{\eps}{6}}[X|Y] & \geq \max _{y \in \cY}\log|\Supp (q(X|Y=y))|,\\
H_{0}^{\frac{\eps}{6}}[Y|X] & \geq \max _{x \in \cX}\log|\Supp (q(Y|X=x))|.
\end{align*}
\end{lemma}
\begin{proof}
Let $q^{(1)}$, $q^{(2)}$ and $q^{(3)}$ be such that
\begin{align*}
H_{0}^{\frac{\eps}{6}}[X, Y] & = \log|\mbox{Supp} (q^{(1)}(X, Y))|,\\
H_{0}^{\frac{\eps}{6}}[X|Y] & = \max _{y \in \cY}\log|\mbox{Supp} (q^{(2)}(X|Y=y))|,\\
H_{0}^{\frac{\eps}{6}}[Y|X] & = \max _{x \in \cX}\log|\mbox{Supp}(q^{(3)}(Y|X=x))|.
\end{align*}
Consider the set,
\beq
E := \mbox{Supp}(q^{(1)}) \cap  \mbox{Supp}(q^{(2)}) \cap  \mbox{Supp}(q^{(3)}). \nonumber
\enq
Furthermore, let us define a positive function $q$ such that
\begin{equation}
\label{optq}
q(X,Y) =
\begin{cases}
p_{XY}(x,y) & \mbox{if } (x,y) \in E, \\
 0    & \mbox{otherwise}.
\end{cases}
\end{equation}
We now show that 
\beq
\sum_{(x,y) \in \cX \times \cY} q(X,Y)\geq 1-\eps /2. \nonumber
\enq
Towards this consider the following set of inequalities
\begin{align*}
\sum_{(x,y) \in \cX \times \cY} q(X,Y)& \overset {a}= \Pr\{E\}\\
& \geq 1 - \Pr\{\mbox{Supp}^c(q^{(1)})\} - \Pr\{\mbox{Supp}^c(q^{(2)})\}\\
& \hspace{5mm}- \Pr\{\mbox{Supp}^c(q^{(3)})\} \\
& \overset{b} \geq 1- \eps/2,
\end{align*}
where $a$ follows from \eqref{optq} and $b$ follows from the fact that for every $i\in \{1, 2, 3\}$, $q^{(i)}(x,y) = p_{XY}(x,y)$ for every $(x,y) \in \mbox{Supp}(q^{(i)})$ and $\sum_{(x,y)\in\cX\times\cY}(q^{(i)})(x,y) \geq 1-\frac{\eps}{6}$. For more details on this see the discussion after Definition \ref{smoothdef1} and Definition \ref{smcond}. The lemma now easily follows from Definition \ref{smcond} and by noticing the fact that
\beq
\mbox{Supp}(q) \subseteq \mbox{Supp}(q^{(i)})~ \forall i\in \{1,2,3\}. \nonumber
\enq

 This completes the proof.
\end{proof}

\subsection{Proof of Theorem \ref{thcsw}}
All three inequalities follow from the result of Renner and Wolf \cite[Theorem 2]{renner-wolf-2005}. For the first,  consider the two party communication solved by Alice on one side and (Bob, Charlie) on the other side. For the second, consider the two party communication solved by Bob on one side and (Alice, Charlie) on the other side. For the third, consider the two party communication solved by (Alice , Charlie) on one side and Bob on the other side. 

\subsection{Relation to previous works}
We now show that the bounds derived in Theorem \ref{thasw} and Theorem
\ref{thcsw} yield the result of Slepian and Wolf
\cite{slepian-wolf-1973} in the iid setting and the result of
Miyake and Kanya \cite[Theorem 1]{miyakaye-kanaya-1995} in the
non-i.i.d. setting. In fact, in \cite{miyakaye-kanaya-1995} Miyake and
Kanaya argued that their result implies the result of Slepian and Wolf
\cite{slepian-wolf-1973} which we restated as Theorem
\ref{slepian-wolf}. Therefore, we will only show that our results
imply the result of Miyake and Kanya \cite[Theorem 1]{miyakaye-kanaya-1995}.

We will state our claim using the following definition.
\begin{definition}(\cite{han})
\label{swdef}
A rate pair $(R_1, R_2)$ is $\eps$-achievable if there exists a pair
of encoding functions $e^{(n)}_{A}$ and $e^{(n)}_{B},$ where
$e^{(n)}_A: \cX^n \to [1: 2^{\ell_A^{(n)}}]$ and $e^{(n)}_B: \cY^n \to
[1: 2^{\ell^{(n)}_B}]$, and a decoder $d^{(n)}_C : [1:
  2^{\ell_A^{(n)}}]\times [1: 2^{\ell_B^{(n)}}] \to \cX^n \times
\cY^n$ such that
\begin{align*}
R_1 &\geq \limsup_{n \to \infty}\frac{\ell_A^{(n)}}{n}\\
R_2 &\geq \limsup_{n \to \infty}\frac{\ell_B^{(n)}}{n},
\end{align*}
and $\lim_{n \to \infty} \Pr\left\{\left(X^n,Y^n\right) \neq
d^{(n)}_C\left(e^{(n)}_A (X^n),e^{(n)}_B(Y^n)\right)\right\} \leq
\eps$.  We say that $(R_1,R_2)$ is achievable if it is $0$-achievable.
\end{definition}
Using this terminology, we have the following theorem.
\begin{theorem}
\label{mksw}
Let $(\bX, \bY):=\left\{X^n,Y^n\right\}_{n=1}^{\infty}$ be a sequence
of arbitrarily distributed random sequence where for every $n$,
$(X^n,Y^n) \sim p_{X^nY^n}$. Fix a pair $(R_1,R_2)$ and $\eps>0$ such that
\begin{align}
R_1&> \limsup_{n \to \infty}\frac{H_{0}^{\frac{\eps}{6}}[X^n|Y^n]}{n} \label{ineq:suff1}\\
R_2&> \limsup_{n \to \infty}\frac{H_{0}^{\frac{\eps}{6}}[Y^n|X^n]}{n} \label{ineq:suff2}\\
R_1 + R_2 &> \limsup_{n \to \infty} \frac{H_{0}^{\frac{\eps}{6}}[X^nY^n]}{n}
\label{ineq:suff12}
\end{align}
Then, $(R_1,R_2)$ is $\eps$-achievable.

If $(R_1,R_2)$ is $\eps$-achievable, then for all $\delta > 0$
\begin{align}
R_1&\geq  \limsup_{n \to \infty} \frac{H_{0}^{\eps + \delta}[X^n|Y^n]}{n} \label{ineq:nec1}\\
R_2&\geq  \limsup_{n \to \infty} \frac{H_{0}^{\eps + \delta}[Y^n|X^n]}{n} \label{ineq:nec2}\\
R_1 + R_2 &\geq  \limsup_{n \to \infty} \frac{H_{0}^{\eps + \delta}[X^nY^n]}{n}. \label{ineq:nec12}
\end{align}
\end{theorem}
\begin{proof}
First consider the achievability result in part (a). Fix a rate pair
$(R_1,R_2)$ satisfying (\ref{ineq:suff1})--(\ref{ineq:suff12}). Consider
the sequence $\left\{\ell_A^{(n)},\ell_B^{(n)}\right\}_{n=1}^{\infty}$, where
$\ell_A^{(n)}:= \floor{n R_1}$ and $\ell_B^{(n)}:=\floor{n R_2}$. Then,
for all large $n$, we have
\begin{align*}
\ell_A^{(n)}&\geq {H_{0}^{\frac{\eps}{6}}[X^n|Y^n]} - \log\left(\frac{\eps}{6}\right)\\
\ell_B^{(n)}&\geq {H_{0}^{\frac{\eps}{6}}[Y^n|X^n]}- \log\left(\frac{\eps}{6}\right)\\
\ell_A^{(n)}+\ell_B^{(n)}&\geq {H_{0}^{\frac{\eps}{6}}[X^nY^n]} - \log\left(\frac{\eps}{6}\right).
\end{align*}
It thus follows from Theorem \ref{thasw} that for all large $n$, there
exists a pair of encoding functions $e_A^{(n)}$ and $e_B^{(n)}$ and a
decoding function $d_C^{(n)}$ such that
$\Pr\left\{\left(X^n,Y^n\right) \neq d^{(n)}_C\left(e^{(n)}_A
(X^n),e^{(n)}_B(Y^n)\right)\right\} \leq \eps$. Thus, from Definition
\ref{swdef} it now follows that the rate pair $(R_1,R_2)$ is
$\eps$-achievable.

Next, to show the converse (that is, the necessity of conditions
(\ref{ineq:nec1})--(\ref{ineq:nec12}), assume that there exist encoding
functions $e^{(n)}_A$ and $e^{(n)}_B$ and decoding functions
$d^{(n)}_C$ such that for all large $n$, we have $nR_1\geq
\ell_A^{(n)}$, $nR_2 \geq \ell_B^{(n)}$ and
$\Pr\left\{\left(X^n,Y^n\right) \neq d^{(n)}_C\left(e^{(n)}_A
(X^n),e^{(n)}_B(Y^n)\right)\right\} \leq \eps + \delta$.
From Theorem \ref{thcsw} we conclude that for all such $n$:
\begin{align*}
nR_1 &\geq \ell^{(n)}_A \geq {H_{0}^{\eps+\delta}[X^n|Y^n]}\\
nR_2 &\geq \ell^{(n)}_B \geq {H_{0}^{\eps+\delta}[Y^n|X^n]}\\
n(R_1+R_2) &\geq  \ell^{(n)}_A+\ell^{(n)}_B \geq {H_{0}^{\eps+\delta}[X^nY^n]}.
\end{align*}
Our claim follows by dividing these equations by $n$ and taking {\em
  limsup} of both sides as $n\rightarrow \infty$.
\end{proof}
We will now show that Theorem \ref{mksw} implies the result of Miyake and
Kanya~\cite[Theorem 2]{miyakaye-kanaya-1995}. Towards this we will use the following convergence result.
\begin{lemma} (Datta and Renner \cite{datta-renner-2009})
\label{assymcond134}
Let $(\bX, \bY) = \{(X^n,Y^n)\}_{n=1}^{\infty}$ be an arbitrary random  sequence taking values over the set $\{(\cX\times\cY)^n\}_{n=1}^{\infty}$,
where $(\cX\times\cY)^n$ is the $n\mhyphen$th Cartesian product of 
$\cX\times \cY$. Then,
\beq
\lim_{\eps\to 0}\limsup_{n\to\infty} \frac{H_{0}^\eps[X^n|Y^n]}{n} = \oH[\bX|\bY], \nonumber
\enq
where $\oH(\bX|\bY)$ is called the spectral-sup conditional entropy
rate of $\bX$ given $\bY$ \cite{han}. In particular, if $(\bX, \bY) =
\{(X^n,Y^n)\}_{n=1}^{\infty}$ is a random sequence of independent and
identically distributed random pairs distributed according to $p_{XY}$
then
\beq
\lim_{\eps\to 0}\lim_{n\to\infty} \frac{H_{0}^\eps[X^n|Y^n]}{n} = H[X|Y].\nonumber
\enq
\end{lemma}
\begin{theorem}
The rate pair $(R_1, R_2)$ is achievable if and only if
\begin{align*}
R_1 &\geq \oH[\bX|\bY]\\
R_2 &\geq\oH[\bY|\bX]\\
R_1+R_2 & \geq \oH[\bX\bY].
\end{align*}
Note that this part is just a restatement of the result of Miyake and
Kanya~\cite{miyakaye-kanaya-1995}.
\end{theorem}
\begin{proof}
To establish the ``if'' direction, it is enough to
show that for all $(R_1, R_2)$ such that
\begin{align*}
R_1 &> \oH[\bX|\bY]\\
R_2 &> \oH[\bY|\bX]\\
R_1+R_2 &> \oH[\bX\bY].
\end{align*}
and all $\eps >0$, $(R_1,R_2)$ is $\eps$-achievable for all large $n$.
Fix $\eps > 0$. Since $H_{0}^{\frac{\eps}{6}}$ is monotone non-increasing in
$\eps$, we conclude from Lemma~\ref{assymcond134} that for all large
$n$,  inequalities (\ref{ineq:suff1})--(\ref{ineq:suff12}) are satisfied.
Our claim follows immediately from part (a). 

For the ``only if'' direction fix a rate pair $(R_1,R_2)$ and a pair
of encoding functions $e^{(n)}_{A}$ and $e^{(n)}_{B},$ where
$e^{(n)}_A: \cX^n \to [1: 2^{\ell_A^{(n)}}]$ and $e^{(n)}_B: \cY^n \to
[1: 2^{\ell^{(n)}_B}]$, and a decoding function $d^{(n)}_C : [1:
  2^{\ell_A^{(n)}}]\times [1: 2^{\ell_B^{(n)}}] \to \cX^n \times
\cY^n$, satisfying Definition~\ref{swdef} with $\eps=0$. Fix
$\gamma>0$, and using the necessary condition in part (a), with
$\eps\leftarrow \frac{\gamma}{2}$ and $\delta \leftarrow
\frac{\gamma}{2}$, conclude that
\begin{align}
R_1&\geq  \limsup_{n \to \infty} \frac{H_{0}^{\gamma}[X^n|Y^n]}{n}\\
R_2&\geq  \limsup_{n \to \infty} \frac{H_{0}^{\gamma}[Y^n|X^n]}{n}\\ 
R_1 + R_2 &\geq  \limsup_{n \to \infty} \frac{H_{0}^{\gamma}[X^nY^n]}{n}. 
\end{align}
Our claim follows by taking the taking limits as
$\gamma$ goes to $0$, and using Lemma~\ref{assymcond134}.
\end{proof}

\section{One-shot source coding with coded side information available at the decoder}
\subsection{Proof of Theorem \ref{oneshotsideinf}}
The techniques used in the proof here are motivated by \cite[Lemma 4.3]{{wyner-1975}} and \cite[Lemma 4]{Kuzuoka-2012}. Let $q \in \cB^{\eps_{A}}(p_{UX})$ and  $\phi \in \cB^{\bar{\eps}_{B}}(p_{UY})$ be such that 
\begin{align}
\label{consm786}
H^{\eps_{A}}_{0}&[X|U]= \log \max_{u \in \cal{U}} |\mbox{Supp}(q(X|U=u))|
\end{align}
and
\begin{align}
\label{smcon12}
I^{\bar{\eps}_{B}}_{\infty}[U;Y] 
& = \log \max_{(u,y) : p_{UY}(u,y)>0}\frac{\phi(u,y)}{p_{U}(u)p_{Y}(y)},
\end{align}
The existence of $\phi$ which satisfies \eqref{smcon12} follows from the fact that the set $\cB^{\bar{\eps}_B}(p_{UY})$ is compact. For every $(u,y) \in \mathcal{U}\times \cY$, let
\beq
\label{fundef}
g(u,y):= \sum_{(u,x)\notin  \mbox{Supp}(q)}p_{X|Y}(x|y).
\enq
Define the following set
\beq
\label{consset2345}
\mathcal{F} := \left\{(u,y)\in \mathcal{U} \times \cY : g(u,y) \leq \eps_{A}^{\frac{1}{2}} \right\}.
\enq

{\bf{Random code generation:}}  Randomly and independently assign  an index $i \in [1: 2^{\ell_{A}}]$ to every realization $x \in \cX$. The realizations with the same index $i$ form a bin $\cB (i)$. Randomly and independently generate $2^{\ell_{B}}$ realizations $u(k)$, $k \in [1:2^{\ell_{B}}]$, each according to $p_U$. Let $\overline{\cU}:= \left\{u(1), u(2), \cdots, u(2^{\ell_B})\right\}$.

{\bf{Encoding:}} If Alice observes a realization $x \in \cB (i)$, then she transmits $i$. For every realization $y \in \cY$ Bob finds an index $k$ such that $(u(k),y) \in \mathcal{F}$. For the case when there are more than one such indices, it sends the smallest one among them. If there is none, he then sends $k =1$.

{\bf{Decoding:}} Charlie finds the unique $x^{\prime} \in \cB(i)$ such that $(x^{\prime},u(k)) \in \mbox{Supp}(q)$. If there is none or more than one, then he declares an error.

{\bf{Probability of error:}} We now calculate probability of error averaged over all code books. 
 Let $M_1$ and $M_2$ be the random chosen indices for encoding $X$ and $Y$. The error in the above mentioned encoding decoding strategy occurs only if one or more of the following error events occur\begin{align*}
E_{1} &= \left\{(U(m_2),Y) \notin \mathcal{F},~ \forall m_2 \in \left[1:2^{\ell_{B}}\right]\right\}\\
E_2 &= \left\{(X, U(M_2)) \notin \mbox{Supp}(q) \right\}\\
E_3 &=  \left\{\exists x^\prime \in \cB(M_1): (x^\prime, U(M_2)) \in \mbox{Supp}(q), x^\prime \neq X\right\}.
\end{align*}
The probability of error is upper bounded as follows
\beq
\label{errana123}
\Pr\{E\} \leq \Pr\{E_1\} + \Pr\{E^c_1 \cap E_2\} + \Pr\{E_3|X\in\cB(1)\}. 
\enq
We now calculate $\Pr\{E_1\}$ as follows
\begin{align}
\Pr\{E_1\} &= \sum_{y \in \cY}p_{Y}(y)\Pr\bigg\{(U(m_2),y) \notin \mathcal{F}, \forall m_2 \in \left[1:2^{\ell_B}\right]\bigg\}\nonumber\\
&\overset{a}= \sum_{y \in \cY}p_{Y}(y) \left(1-\sum_{u:(u,y) \in \mathcal{F}}P_{U}(u)\right)^{2^{\ell_B}}\nonumber\\
&\overset{b} \leq\sum_{y \in \cY}p_{Y}(y) \bigg(1-2^{-I^{\bar{\eps}_{B}}_{\infty}[U;Y]}\sum_{u:(u,y) \in \mathcal{F}}\frac{\phi(u,y)}{p_Y(y)}\bigg)^{2^{\ell_B}}\nonumber\\
&\overset{c}\leq\sum_{y \in \cY}p_{Y}(y)e^{\left(-2^{\ell_B}2^{-I^{\bar{\eps}_{B}}_{\infty}[U;Y]}\sum_{u:(u,y) \in \mathcal{F}}\frac{\phi(u,y)}{p_Y(y)}\right)}\\
& \overset{d} \leq 1-\sum_{(u,y) \in \mathcal{F}} \phi(u,y) + e^{-2^{\ell_B}2^{-I^{\bar{\eps}_{B}}_{\infty}[U;Y]}} \nonumber \\
& \overset{e} \leq \bar{\eps}_{B} + \Pr\{(U,Y) \notin \mathcal{F}\} + e^{-2^{\ell_B}2^{-I^{\bar{\eps}_{B}}_{\infty}[U;Y]}}, \nonumber
\end{align}
where $a$ follows because $U(1),\dots,U(2^{\ell_B})$ are independent and subject to identical distribution $p_U$; $b$ follows because for every $(u,y) \in \mathcal{U} \times\cY$ we have 
\beq
\label{justifyb}
p_{U}(u)\geq 2^{-I^{\bar{\eps}_{B}}_{\infty}[U;Y]}\frac{\phi(u,y)}{p_Y(y)},
\enq
where \eqref{justifyb} follows from \eqref{smcon12}; $c$ follows from the inequality $(1-x)^y \leq e^{-xy} ~ (0 \leq x \leq 1, y\geq 0)$ in our setting $x = 2^{-I^{\bar{\eps}_{B}}_{\infty}[U;Y]}\sum_{u:(u,y) \in \mathcal{F}}\frac{\phi(u,y)}{p_Y(y)}$ and $y =2^{\ell_B}$; $d$ follows because of the inequality $e^{-xy} \leq 1-x+e^{-y} ~(0\leq x \leq 1, y \geq 0)$ in our setting $x = \sum_{u :(u,y)\in \mathcal{F}} \frac{\phi(u,y)}{p_Y(y)}$ and $y =2^{\ell_B}2^{-I^{\bar{\eps}_{B}}_{\infty}[U;Y]}$ and $e$ is true because of the following arguments
 \begin{align}
 1-\bar{\eps}_{B} & \overset {a}\leq \sum_{(u,y) \in \mathcal{U} \times \cY} \phi(u,y) \nonumber\\
 & = \sum_{(u,y) \in \mathcal{F}^c} \phi(u,y)+  \sum_{(u,y) \in \mathcal{F}}\phi(u,y)\nonumber\\
 & \overset{b} \leq \Pr\{\mathcal{F}^c\} +  \sum_{(u,y) \in \mathcal{F}}\phi(u,y)\nonumber\\
 \label{rearrange4}
 &  \leq \Pr\{(U,Y) \notin \mathcal{F}\} +\sum_{(u,y) \in \mathcal{F}}\phi(u,y),
 \end{align}
 where $a$ and $b$ both follow from the fact that $\phi(u,y)
  \in \cB^{\bar{\eps}_{B}}(p_{UY})$. By rearranging the terms in \eqref{rearrange4} we get
 \beq
 \label{errb12}
 1- \sum_{(u,y) \in \mathcal{F}}\phi(u,y) \leq \bar{\eps}_{B}+\Pr\{(U,Y) \notin \mathcal{F}\}. 
 \enq
We now calculate $\Pr\{(U,Y)\notin\mathcal{F}\}$ as follows 
\begin{align}
\Pr\{(U,Y) \notin \mathcal{F}\} &= \Pr\{g(U,Y) \geq \eps_{A}^{\frac{1}{2}}\} \nonumber\\
&\overset{a} \leq \eps^{-\frac{1}{2}}_{A} \mathbb{E}_{UY}(g(U,Y)) \nonumber\\
& \leq \eps_{A}^{-\frac{1}{2}} \sum_{(u,y) \in \mathcal{U }\times \cY}p_{UY}(u,y) g(u,y) \nonumber\\
&\overset{b} = \eps_{A}^{-\frac{1}{2}} \sum_{(u,y) \in \mathcal{U}\times \cY}p_{UY}(u,y)\sum_{x : (u,x) \notin \mbox{Supp}(q)}p_{X|Y}(x|y) \nonumber\\
& = \eps^{-\frac{1}{2}}_{A}\sum_{\substack{(x,u,y) \in \cX\times \mathcal{U }\times \cY\\ (u,x) \notin \mbox{Supp}(q)}}p_{XUY}(x,u,y) \nonumber\\
& = \eps^{-\frac{1}{2}}_{A} \sum_{(u,x) \notin \mbox{Supp}(q)}p_{UX}(u,x) \nonumber\\
\label{finc3}
&\overset{c} \leq \eps^{\frac{1}{2}}_{A},
\end{align}
where $a$ follows from Markov's inequality; $b$ follows from \eqref{fundef} and $c$ follows because of the following arguments
\begin{align}
1-\eps_{A} & \overset {a}\leq \sum_{(u,x)\in \mathcal{U} \times \cX}q(u,x)\nonumber\\
\label{rearrang5}
&\overset{b}\leq \sum_{(u,x) \in \mbox{Supp}(q)}p_{UX}(u,x),
\end{align}
where $a$ follows from \eqref{consm786} and $b$ follows because $q(u,x)\leq p_{UX}(u,x)$, for every $(u,x) \in \mathcal{U} \times \cX$. By rearranging the terms in \eqref{rearrang5} we get
\beq
1-\sum_{(u,x) \in \mbox{Supp}(q)}p_{UX}(u,x) \leq \eps_{A}. \nonumber
\enq
The second term in \eqref{errana123} is calculated as follows
\begin{align}
\Pr\{E^c_1\cap E_2\} & \leq \Pr\{E_2 \mid E^c_1\} \nonumber\\
&\leq \sum_{\substack{(x,u,y)\in \cX\times\cU\times\cY\\(u,y) \in \mathcal{F}, (u,x)\notin \mbox{Supp}(q)}}\Pr\left\{ Y=y, U(M_2)= u\mid E^c_1\right\}\Pr\left\{X=x \mid Y=y, U(M_2) = u, E^c_1\right\} \nonumber\\
&\overset{a}=\sum_{\substack{(u,y)\in \cU\times\cY\\(u,y) \in \mathcal{F}}}\Pr\left\{ Y=y, U(M_2) = u\mid E^c_1\right\}\sum_{x : (u,x) \notin \mbox{Supp}(q)}p_{X\mid Y}(x \mid y)\nonumber\\
\label{finc2}
&\overset{b}\leq \eps_{A}^{\frac{1}{2}},
\end{align}
where $a$ follows because $E^c_1$ and $M_2$ are completely determined based on ~$\overline{\cU}$, $Y$ and $\overline{\cU}$ was generated independent of $(X,Y)$ and $b$ follows from \eqref{consset2345}. From \eqref{errb12}, \eqref{finc3} and \eqref{finc2}  it follows that 
\begin{align}
\Pr\{E_1\}+\Pr\{E^c_1 \cap E_2\} \leq \bar{\eps}_{B} + 2\eps_{A}^{\frac{1}{2}}+ e^{-2^{\ell_B}2^{-I^{\bar{\eps}_{B}}_{\infty}[U;Y]}}. \nonumber
\end{align}
Thus, from the choice of $\ell_B$ and $\eps_A$ and $\bar{\eps}_B$ as mentioned in the theorem, it now follows that
\beq
\label{finc4}
\Pr\{E_1\}+\Pr\{E^c_1 \cap E_2\} \leq \eps_B.
\enq
Finally, the third term in \eqref{errana123} is calculated as follows 
\begin{align}
\Pr\{E_3\} &= \Pr\{E_3|\cB(1)\}\nonumber\\
& = \sum_{(x,u) \in \cX \times \mathcal{U}} \Pr\big\{(X,U) = (x,u)|X \in \cB(1)\big\} \Pr \bigg\{\exists x^\prime \neq x : x^\prime \in \cB(1)\mbox{and} ~q(x^\prime,u)>0\nonumber \\
 &\hspace{82mm}\big| x \in \cB(1),(X,U) = (x,u)\bigg\} \nonumber\\
& \leq \sum_{(x,u) \in \cX \times \mathcal{U}}p_{XU}(x,u) \sum_{\substack {x^\prime \neq x \\ q(x^\prime,u) > 0}}\Pr \{x^\prime \in \cB(1)\} \nonumber\\
&\leq  2^{-\ell_A}\sum_{(x,u) \in \cX \times \mathcal{U}}p_{XU}(x,u)  \max_{u \in \mathcal{U}} \sum_{x^\prime:q(x^\prime,u)>0} 1 \nonumber\\
& \overset{a} = 2^{-\ell_A}\sum_{(x,u) \in \cX \times \mathcal{U}}p_{XU}(x,u) \max_{u \in \mathcal{U}} |\mbox{Supp}(q(X|U=u))|\nonumber\\
\label{sw123}
& = 2^{-\ell_A}\max_{u \in \mathcal{U}} |\mbox{Supp}(q(X|U=u))|, 
\end{align}
where $a$ follows because $q(X=x|U=u): = \frac{q(x,u)}{p_U(u)}$ and $q(X=x|U=u): =0$ if $p_{U}(u) = 0$.
Thus from \eqref{finc4} and \eqref{sw123} it follows that
\beq
\Pr\{E\} \leq \eps_B + 2^{-\ell_A}\max_{u \in \mathcal{U}} |\mbox{Supp}(q(X|U=u))|. \nonumber
\enq
Thus, from the choice of $\ell_A$ and $\eps_B$ as mentioned in the theorem it now follows that
$$\Pr\{E\} \leq \eps.$$
This completes the proof.
\subsection{Alternate achievability bounds}
In this section, we analyse the helper problem using different techniques and obtain a different version of Theorem \ref{oneshotsideinf}. Note that the final error probability depends linearly on $\eps_A$; in Theorem \ref{oneshotsideinf}, this dependence was of the form $\sqrt{\eps_A}$. Also, the  proof techniques used in deriving these alternate achievability bounds are  completely different from the techniques used in the proof of Theorem 6.

\subsection*{Part I: Alice's message to Charlie}

Fix the random variable $U$ and other quantities as in the assumption
of the theorem.  The proof has two parts. First, we observe that if
Charlie is in possession of $U$, then Alice need send only about
$H_{0}^{\varepsilon_A}(X|U)- \log(\eps_A)$ bits to ensure that Charlie can
reconstruct $X$ with probability $2\eps_A$. Indeed, we have the following
result.
\begin{theorem}(Renner and Wolf \cite{renner-wolf-2005}) \label{thm:RennerWolf}
Let $(X,U)$ be random variables taking values in the set ${\mathcal X}
\times {\mathcal U}$ and let $\ell_A \geq H_0^{\varepsilon_A}[X \mid U] + \log (1/\eps_A)$, where $2\eps_A \leq \eps$.
Then, there is an encoding function $e_A: {\mathcal X} \rightarrow
[1:2^{\ell_A}]$ and a decoding function $d_{AC}: [1:2^{\ell_A}]\times \cU
\rightarrow {\mathcal X}$ such that
\beq
\label{rennertheorem}
\Pr\left\{d_{AC}(e_{A}(X),U) \neq X\right\} \leq \eps.
\enq
\end{theorem}
Thus, if inequality in \eqref{altb1} holds and $U$ is available with Charlie then, he can reconstruct $X$ with probability of error at most $2\eps_A$.
Furthermore, in Theorem \ref{thm:RennerWolf} if Alice is in possession of a random variable ${U^\prime}$ instead
of $U$ (that is, the pair $(X,{U^\prime})$ takes values in ${\mathcal X}\times {\mathcal U}$)  . Then, from \eqref{rennertheorem} it follows that 
\[ \Pr[d_{AC}(e_{A}(X),{U^\prime}) \neq X] \leq 2\eps_A+ \frac{1}{2}{\|p_{XU}-p_{XU^\prime}\|}.\]
In light of this, it is enough to devise a means by which $U$ (or some
close approximation ${U^\prime}$) can be provided to Charlie. We will
show that Bob can accomplish this whenever $\ell_B$ is sufficiently
large so that \eqref{altb2} is satisfied.

\subsection*{Part II: Bob helps Charlie generate $U$}
It will be convenient to work in a setting where Bob and Charlie have
access to shared randomness (generated independently of the inputs
$(X,Y)$). In the end, the shared randomness will be set to a fixed
value, and eliminated.  Furthermore, instead of the Markov Chain
$X$---$Y$---$U$ we will work with $X$---$Y$---$U'$, where $U'$ is
defined in the following lemma. 
\begin{lemma}
\label{prop:auxU}
Let $\phi \in \cB^{\varepsilon}(p_{UY})$ be such
that
\[ I_\infty^\varepsilon[U;Y] = \max_{(u,y): \phi_{UY}(u,y) > 0} \log \frac{\phi(u,y)}{p_U(u) p_Y(y)} .\]
For $y \in {\mathcal Y}$, let $\alpha_y := 1- 
\frac{\sum_u\phi(u,y)}{p_Y(y)}$.  Let
\beq
\label{lemmares}
p_{U^\prime|Y=y}(u)  =
\begin{cases}
\frac{\phi(u,y)}{p_Y(y)(1-\alpha_y)} & \mbox{if }~ \alpha_y \leq \frac{1}{2}; \\
p_{U}(u)        & \mbox{otherwise.}
\end{cases}
\enq
Then,
\begin{align}
\label{claim1234}
\|p_{UY} - p_{U'Y} \| &\leq 6 \varepsilon;\\ 
\label{inf1235}
D_\infty(p_{U^\prime Y} \| p_{U} \times p_{Y}) &\leq
\max\left\{0, I_\infty^\varepsilon[U;Y] + 1\right\}.
\end{align}
\end{lemma}
\begin{proof}
We first establish \eqref{claim1234}. Let $\cA:= \{y \in \cY : \alpha_y > \frac{1}{2}\}$. Then,
\begin{align}
\|p_{UY}-p_{U^\prime Y}\|&= \sum_{(u,y) \in \cU \times \cY}\big|p_{UY}(u,y) - p_{U^\prime Y}(u,y)\big| \nonumber\\
\label{trianginter}
&= \sum_{y\in \cA}\sum_{u \in \cU}\big|p_{UY}(u,y) - p_{U^\prime Y}(u,y)\big|+\sum_{y \in \cA^c}\sum_{u \in \cU}\big|p_{UY}(u,y) - p_{U^\prime Y}(u,y)\big| 
\end{align}
Let us now bound each of the terms on the R.H.S. of \eqref{trianginter}. The first term on the R.H.S. of \eqref{trianginter} is bounded as follows:
\begin{align}
\sum_{y\in \cA}\sum_{u \in \cU}\big|p_{UY}(u,y) - p_{U^\prime Y}(u,y)\big|&\overset{a} = \sum_{y\in \cA}p_{Y}(y)\sum_{u \in \cU}\big|p_{U|Y}(u \mid y)-p_{U^\prime \mid Y}(u \mid y)\nonumber\big| \nonumber\\ 
&\leq 2\Pr\left\{\alpha_Y > \frac{1}{2}\right\} \nonumber\\
\label{bb2}
&\overset{b}\leq 4\eps,
\end{align}
where $a$ follows from the fact that the distance between any two distributions can be at most $2$ and $b$ follows from Markov's inequality and the fact that $\mathbb{E}_Y\left[\alpha_Y\right]= 1-\mathbb{E}_Y\left[\frac{\sum_u\phi(u,Y)}{p_Y}\right]  \leq \eps$. The second term on the R.H.S. of \eqref{trianginter} is bounded as follows:
\begin{align}
 \sum_{y \in \cA^c}\sum_{u \in \cU}\big|p_{UY}(u,y) - p_{U^\prime Y}(u,y)\big| &= \sum_{y \in \cA^c}\sum_{u \in \cU}\bigg|p_{UY}(u,y)- \phi(u,y)+ \phi(u,y)- \frac{\phi(u,y)}{1-\alpha_y}\bigg| \nonumber\\
&\leq\sum_{y\in\cA^c}\sum_{u\in\cU}\big|p_{UY}(u,y)- \phi(u,y)\big|+ \sum_{y\in\cA^c}\sum_{u\in\cU}\bigg|\phi(u,y)- \frac{\phi(u,y)}{1-\alpha_y}\bigg|\nonumber\\
&\overset{a}\leq \eps +\sum_{y\in\cA^c}\sum_{u\in\cU}\bigg|\phi(u,y)- \frac{\phi(u,y)}{1-\alpha_y}\bigg|\nonumber\\
&\leq \eps + \sum_{y\in\cA^c}\sum_{u\in\cU}\frac{\alpha_y}{1-\alpha_y}\phi(u,y),\nonumber\\
&= \eps+\sum_{y\in\cA^c}\frac{\alpha_y}{1-\alpha_y}p_{Y}(y)\frac{\sum_{u\in\cU}\phi(u,y)}{p_{Y}(y)}\nonumber\\
&\overset{b}\leq \eps+ \sum_{y\in\cA^c}p_{Y}(y)\alpha_y\nonumber\\
&\overset{c}\leq 2\eps,
\label{cla}
\end{align}
where $a$ follows from the fact that $\phi \in \cB^{\eps}(p_{UY})$; $b$ follows because $\alpha_y := 1- \frac{\sum_u\phi(u,y)}{p_Y(y)}$ and $c$ follows because $\mathbb{E}_Y\left[\alpha_Y\right]\leq \eps$. 

We now prove \eqref{inf1235}. From the definition of $P_{U^\prime Y}$ it is easy to see that for every $(u,y)$ we have 
\beq
\label{uprimeDinft}
\log\frac{p_{U^\prime|Y}(u|y)}{p_{U}(u)} =
\begin{cases}
\log\frac{\phi(u,y)}{p_{U}(u)p_{Y}(y)}+1 & \mbox{if }~ \alpha_y \leq \frac{1}{2}; \\
0        & \mbox{otherwise.}
\end{cases}
\enq
The claim in \eqref{inf1235} now immediately follows from \eqref{uprimeDinft} and from the Definition \ref{smoothorderinf}. This completes the proof.
\end{proof}
We also need Lemma \ref{prop:rejectionsampling} (implicit in earlier works;
see Refs. \cite{jain-radhakrishnan-sen-2005}, \cite{jain-radhakrishnan-sen-05}), which bounds the number of bits Bob needs to send to Charlie so that Charlie can simulate the random variable in Bob's possession.  
\begin{lemma}\label{prop:rejectionsampling}
Let $U \sim p_U$ take values over the set ${\mathcal U}$.  Let $L$
be a positive integer. Let $R$ be distributed uniformly in the real unit
cube $[0,1]^{L+1}$.  Then, there is a decoding function $d: [0,1]^{L+1}
\times [1:L] \rightarrow {\mathcal U}$ such that for every distribution
$q$ on ${\mathcal U}$, there is an encoder $e_q: [0,1]^{L+1}
\rightarrow [1:L]$, such that the random variable $V= d(R,e_q(R))$ (notice the dependence of $V$ on $q$)
satisfies $\|p_{V} - q\| \leq 2\exp\left(- L2^{-D_{\infty}(q \|
p_U)}\right)$.
\end{lemma}
\begin{proof}
Let $U$ be as in the lemma, let $Z$ be uniformly distributed
over $[0,1]$. We will use the following rejection-sampling idea to
generate a random variable $\hat{U}$ with distribution $q$ from $U$. For $u
\in \mathcal{U}$ and $z \in [0,1]$, let $b: \mathcal{U} \times [0,1]
\rightarrow \{0,1\}$ be defined by $b(u,z)=1$ iff $z \leq 2^{-D_\infty(q
\| p_U)}q(u)/p_U(u)$. Let $B= b(U,Z)$. We now make the following claims which will help us to prove
the desired result. For now we will assume these claims. However, we will prove them towards the end.
\begin{claim}
\label{c111}
$ \Pr[B=1\mid U=u] = 2^{-D_\infty(q \| p_U)}{q(u)/}{p_U(u)}.$
\end{claim}
\begin{claim}
 \label{cl:two}
 $\Pr[B=1] = 2^{-D_\infty(q \| p_U)}. $
  \end{claim}
 \begin{claim}
  \label{cl:one}
 $
 \Pr[U=u \mid B=1] = q(u).
 $
 \end{claim}
 
 It will be easier to present the construction of the encoder and decoder by referring to Bob and Charlie. Assume Bob posses a random variable which takes values over the set $\cU$ and is distributed according to the distribution $q$, and Charlie wishes to construct a random variable $V$ with distribution close to $q$. Suppose Bob and Charlie share a sequence $T = \big\{ (U_1,Z_1),
(U_2, Z_2), \ldots, (U_L,Z_L) \big\}$ of $L$ i.i.d. pairs each with the
distribution of $(U,Z)$.  Furthermore, for $j=1,2,\ldots, L$, let
$B_j= b(U_j, Z_j)$. Bob's message is constructed as follows. Let
$\mathbf{i}$ be the smallest index $j \in [1:L]$ such that $B_j=1$; if
no such index exists, let $\mathbf{i}=L+1$. By Claim \ref{cl:one}, for
$j \in [1:L]$ and for all $u \in \mathcal{U}$, 
\beq
\label{ui}
\Pr[U_{\mathbf{i}}=u
\mid \mathbf{i}= j]=q(u).\enq On the other hand, 
\begin{align}
\Pr[\mathbf{i}=L+1] &= \left(1-2^{-D_\infty(q || p_U)}\right)^L \nonumber\\
\label{inq}
& \leq \exp\left(- L2^{-D_{\infty}(q \|p_U)}\right),
\end{align}
where \eqref{inq} follows from the inequality $(1-x)^y \leq e^{-xy}$, for $0\leq x \leq 1$ and $y\geq 0$. Bob can thus send $\mathbf{i}$ when it is in the range
$[1:L]$, and send an arbitrary value, say $1$, when
$\mathbf{i}=L+1$. Charlie decodes this as $V=U_{\mathbf{i}}$ (Charlie
has no use for $Z_1,Z_2,\ldots, Z_L$).  Since the sequence
$\overline{U}=U_1,U_2,\ldots, U_L$ takes values in the finite set
$\mathcal{U}^L$, one can realize $\overline{U}$ as function of a
random variable taking values uniformly in $[0,1]$. Consequently,
$T$ may be viewed as function of a random variable $R$ distributed uniformly in $[0,1]^{L+1}$. To conclude the proof we now need to show that 
$\|p_V-q\| \leq 2\exp\left(- L2^{-D_{\infty}(q \|
p_U)}\right)$. Towards this let us first calculate $\Pr[V=u],$ for every $u \in \cU$ as follows:
\begin{align}
\Pr[V=u] &= \sum_{j \in [1:L]}\Pr [\mathbf{i} =j]\Pr[V=u \mid \mathbf{i} =j] + \Pr [\mathbf{i} =L+1]\Pr[V=u \mid \mathbf{i} =L+1] \nonumber\\
&\overset{a}= \sum_{j \in [1:L]}\Pr [\mathbf{i} =j]\Pr[U_{\mathbf{i}}=u \mid \mathbf{i} =j] + \Pr [\mathbf{i} =L+1]\Pr [U_1 = u \mid \mathbf{i} =L+1] \nonumber\\
&\overset{b} = \sum_{j \in [1:L]}\Pr [\mathbf{i} =j]q(u)+\Pr [\mathbf{i} =L+1]\Pr [U_1 = u \mid \mathbf{i} =L+1]  \nonumber\\
& = \left(1-\Pr [\mathbf{i} =L+1]\right)q(u)+\Pr [\mathbf{i} =L+1]\Pr [U_1 = u \mid \mathbf{i} =L+1] \nonumber\\
\label{ffi}
& = q(u) + \bigg(\Pr [U_1 = u \mid \mathbf{i} =L+1] - q(u)\bigg)\Pr [\mathbf{i} =L+1],
\end{align}
where $a$ follows from the setup of the protocol discussed above and $b$ follows from \eqref{ui}. The claim of the lemma now follows from the following set of inequalities
\begin{align}
\|p_V-q\| &= \sum_{u \in \cU} |p_V(u)- q(u)| \nonumber\\
&\overset{a}= \sum_{u \in \cU}\bigg|q(u) + \bigg(\Pr [U_1 = u \mid \mathbf{i} =L+1] - q(u)\bigg)\Pr [\mathbf{i} =L+1]- q(u) \bigg|\nonumber\\
&=\|p_{U_1 \mid \mathbf{i} =L+1} - q\|\Pr [\mathbf{i} =L+1]\nonumber\\
&\overset{b} \leq 2 \Pr [\mathbf{i} =L+1]\nonumber\\
& \overset{c}\leq 2 \exp\left(- L2^{-D_{\infty}(q \|p_U)}\right),
\end{align}
where $a$ follows from \eqref{ffi}, $b$ follows because of the fact that the $L_1$ distance between two distributions can be at most $2$ and $c$ follows from \eqref{inq}. This completes the proof of the lemma.

Let us now prove Claim \ref{c111}, Claim \ref{cl:two} and Claim \ref{cl:one}.

{{Proof of Claim \ref{c111}:}}  $\Pr[B=1|U=u] = \Pr\left\{Z \leq 2^{-D_\infty(q\| p_U)}q(u)/p_U(u) \right\} = 2^{-D_\infty(q \| p_U)}q(u)/p_U(u).$

{{Proof of Claim \ref{cl:two}:}} $\Pr[B=1] = \mathbb{E}_{U}\Pr[B=1\mid U]$. Claim $2$ now follows from Claim \ref{c111}.

{{Proof of Claim \ref{cl:one}:}} 
\begin{align*}
 \Pr[U=u\mid B=1] &= \frac{\Pr[B=1 \mid U = u]}{\Pr[B=1]}\Pr[U=u]\\
 &\overset{a}= \frac{2^{-D_\infty(q \| p_U)}{q(u)/}{p_U(u)}}{2^{-D_\infty(q \| p_U)}}p_{U}(u)\\
 &= q(u),
\end{align*}
where $a$ follows from Claim \ref{c111} and Claim \ref{cl:two}.
\end{proof}

In our application, $R$ is the shared randomness between Bob and
Charlie, $q$ is $p_{U'| Y=y}$ and Bob sends messages of length
$\ell_B=\lceil{\log L}\rceil$. In this setting, we conclude that for every  $\delta \in (0,1)$,
$\|p_{V} - q\| \leq 2\delta$, if $\ell_B \geq D_{\infty}(q \| p_U) + \log\ln(1/\delta)$.

\subsection*{Proof of Theorem \ref{jaikumar's proof}}
Fix random variables $X$, $Y$ and $U$ as in the assumption of the
Theorem \ref{jaikumar's proof}. Let us (temporarily) assume that Bob and Charlie share a
random variable $R$ as in
Lemma~\ref{prop:rejectionsampling}. The encoding and decoding
are performed as follows.  Alice on receiving a realization $x$ of
$X$, sends Charlie a message $M_A=e_A(x)$ of length $\ell_A$, obtained
by applying Theorem~\ref{thm:RennerWolf} with the $X$ and $U$ given in
the theorem; let the corresponding decoding function for Charlie be
$d_{AC}$. Now consider Bob.  Let $U'$ be the random variable that was
derived in Proposition~\ref{prop:auxU}.  Bob, on receiving a
realization $y$ for $Y$, sends Charlie the message $M_B= e_q(R)$,
where $e_q$ is the encoding function guaranteed by
Lemma~\ref{prop:rejectionsampling} for $q=p_{U'|Y=y}$ (see Lemma \ref{prop:auxU} for the definition of $U^\prime$)
and $L=2^{\ell_B}$; let the corresponding decoding function for
Charlie be $d_{BC}$. Charlie's reconstruction of $X$ is the random
variable $X'=d_{AC}(M_A, d_{BC}(M_B,R))$. It remains to establish the
bound on the error claimed in \ref{eq:prob}. Let $V=
d_{BC}(M_B,R)$. Then,
\begin{align}
\Pr\{X \neq X'\} &\leq \Pr\{X \neq d_{AC}(M_A,U)\} + \frac{1}{2} \|p_{XV} - p_{XU}\| \nonumber \\
               &  \leq  2 \eps_A  + \frac{1}{2} \bigg(\|p_{XV}-p_{XU'}\| + \|p_{XU'} - p_{XU} \|\bigg),
\end{align}
where the last inequality above follows from Theorem \ref{thm:RennerWolf}. Since $X$---$Y$---$U$ and $X$---$Y$---$U^\prime$ it follows that $\|p_{XU'} - p_{XU} \| \leq \|p_{YU'} - p_{YU}\|$ and $\|p_{XV} - p_{XU'}\| \leq \|p_{YV} - p_{YU'}\|$. Thus, from from Lemma \ref{prop:auxU} and Lemma \ref{prop:rejectionsampling} we have the following bounds 
\begin{align*}
\|p_{XU'} - p_{XU}\| &\leq 6 \varepsilon_B\\
\|p_{XV} - p_{XU'}\| & \leq 2\eps_B.
\end{align*}
Thus, the claim mentioned in (\ref{eq:prob}) now immediately follows from
this. (Since the error is averaged over shared randomness $R$, there
must be one realization, conditioned on which the error bounds still
hold; thus $R$ is only a temporary device used in the proof, and will
have no role in the final encoding and decoding.)
\subsection {Proof of Theorem \ref{conversesideinf}
}
We shall use the following notation 
\beq
\hat{X} = d_{C}(e_A(X), e_B(Y)).  \nonumber
\enq
We first give the proof for the lower bound on $\ell_A$. The proof follows using ideas similar to that used in the proof of the lower bounds given in \cite[Theorem 1] {sharma-warsi-allerton}. Notice the following set of inequalities
\begin{align*}
\ell_A& \overset{a}{\geq} H_0[e_A(X)] \\
& \overset{b}{\geq} H_0[e_A(X) | U] \\
& \overset{c}{=} H_0[e_A(X), e_B(Y) | U] \\
& \overset{d}{\geq} H_0 \left[d_{C}(e_A(X), e_B(Y)) | U \right]\\
&= H_{0}[\hat{X}|U]\\
& \overset{e}{\geq} H_0^{\eps}[X|U],
\end{align*}
where $a$ follows from the definition of $e_A$; $b$ follows from the fact that conditioning reduces $H_0$; $c$ easily follows from Definition \ref{condzero} and the fact that $U = e_B(Y)$; $d$ follows from the fact that taking function reduces $H_0$; $e$ follows from follows from  Definition \ref{smcond} and from  \cite[Section \rom{4}]{schoe-isit-2007}. We now prove the lower bound on $\ell_B$. Let us define the following set 
\beq
\label{settrick}
\cA := \left\{u \in \cU: p_{U}(u) <\eps 2^{-\ell_B}\right\},
\enq
where $\cU:= [1:2^{\ell_B}]$. It now follows from our construction of the set $\cA$ that $\Pr\{\cA\} < \eps$. Let the function $\phi : \cU \times \cY \to [0,1]$ be defined by
\beq
\label{phiconverse}
\phi(u,y) :=
\begin{cases}
p_{UY}(u,y) & \mbox{if}~ u \notin \cA ,\\ 
0        & \mbox{otherwise}.
\end{cases}
\enq
It easily follows from \eqref{phiconverse} that $\sum_{(u,y)\in \cU\times\cY}\phi(u,y) \geq 1-\eps$. We now have the following set of inequlities
\begin{align*}
I^{\eps}_{\infty}[U;Y] & \overset{a}\leq \max_{(u,y)\in \Supp(\phi)}\log \frac{\phi(u,y)}{p_{U}(u)p_{Y}(y)}\\
& \overset{b} =\max_{(u,y)\in \Supp(\phi)}\log\frac{p_{UY}(u,y)}{p_{U}(u)p_{Y}(y)}\\
&=\max_{(u,y)\in \Supp(\phi)} \log\frac{p_{U\mid Y}(u\mid y)}{p_{U}(u)}\\
& \overset{c} \leq \log\frac{1}{\eps2^{-\ell_B}}\\
& = -\log\eps+ \ell_B,
\end{align*}
 where $a$ follows from Definition \ref{smoothorderinf}; $b$ follows from \eqref{phiconverse} and $c$ follows from \eqref{settrick}. This completes the proof.
\subsection{Relation to previous works}
We now show that the bounds derived in Theorem \ref{oneshotsideinf},
Theorem \ref{jaikumar's proof} and Theorem \ref{conversesideinf} yield
the result of \cite{wyner-1975} in the asymptotic i.i.d. setting and
the result of Miyake and Kanya \cite[Theorem 2]{miyakaye-kanaya-1995}
in the asymptotic non-i.i.d. setting. In fact, in
\cite{miyakaye-kanaya-1995} Miyake and Kanaya argued that their result
implies the result of Wyner which we restated as Theorem
\ref{slepian-wolf}. Therefore, we will only show that our results
imply the result of Miyake and Kanya \cite[Theorem 2]{miyakaye-kanaya-1995}. 

We will state our claim using the following definition.
\begin{definition}(\cite{han})
\label{sidddef}
A rate pair $(R_1, R_2)$ is $\eps$-achievable if and only if there exists a pair of encoding functions $e^{(n)}_{A}$  and $e^{(n)}_{B},$ where 
$e^{(n)}_A: \cX^n \to [1: 2^{\ell_A^{(n)}}]$ and $e^{(n)}_B: \cY^n \to [1: 2^{\ell^{(n)}_B}]$, and a decoder $d^{(n)}_C : [1: 2^{\ell_A^{(n)}}]\times [1: 2^{\ell_B^{(n)}}] \to \cX^n $ such that 
\begin{align*}
R_1 &\geq \limsup_{n \to \infty}\frac{\ell_A^{(n)}}{n}\\
R_2 &\geq \limsup_{n \to \infty}\frac{\ell_B^{(n)}}{n},
\end{align*}
and $\lim_{n \to \infty} \Pr\left\{X^n \neq d^{(n)}_C\left(e^{(n)}_A (X^n),e^{(n)}_B(Y^n)\right)\right\} \leq \eps$. We say that $(R_1,R_2)$ is achievable if it is $0$-achievable
\end{definition}
Using this terminology, we have the following theorem.
\begin{theorem}
\label{mksw}
Let $(\bX, \bY):=\left\{X_n,Y_n\right\}_{n=1}^{\infty}$ be a sequence
of arbitrarily distributed random sequence where for every $n$,
$(X^n,Y^n) \sim p_{X^nY^n}$. 
Fix a pair $(R_1,R_2)$ and $\eps>0$ such that
\begin{align}
R_1&> \limsup_{n \to \infty}\frac{H_{0}^{{\eps}_{A}}[X^n|U^n]}{n} \label{ineq:suffh1}\\
R_2&> \max\left\{0,\limsup_{n \to \infty} \frac{I^{{\eps}_{B}}_{\infty}[U^n;Y^n]}{n}\right\},\label{ineq:suffh2}
\end{align}
where the sequence of random variables $U^n$ satisfies the Markov chain condition $p_{X^nY^nU^n}=p_{X^nY^n}p_{U^n\mid Y^n}$ for each $n$ and  $2\eps_A + 4\eps_B \leq \eps$. Then, $(R_1,R_2)$ is $\eps$-achievable.

If $(R_1,R_2)$ is $\eps$-achievable, then for all $\delta > 0$
\begin{align}
R_1&\geq  \limsup_{n \to \infty} \frac{H_{0}^{\eps + \delta}[X^n|U^n]}{n} \label{ineq:nech1}\\
R_2&\geq \limsup_{n \to \infty}\frac{I^{\eps+\delta}_{\infty}[U^n;Y^n]}{n}\label{ineq:nech2},
\end{align}
where the sequence of random variables $U^n$ satisfies the Markov chain condition $p_{X^nY^nU^n}=p_{X^nY^n}p_{U^n\mid Y^n}$ for each $n$.
\end{theorem}
\begin{proof}
First consider the achievability result in part (a). Fix a rate pair
$(R_1,R_2)$ satisfying (\ref{ineq:suffh1})--(\ref{ineq:suffh2}). Consider
the sequence $\left\{\ell_A^{(n)},\ell_B^{(n)}\right\}_{n=1}^{\infty}$, where
$\ell_A^{(n)}:= \floor{n R_1}$ and $\ell_B^{(n)}:=\floor{n R_1}$. Then,
for all large $n$, we have
\begin{align*}
\ell_A^{(n)}&\geq {H_{0}^{{\eps}_{A}}[X^n|U^n]} +\log\left(1/\eps_A\right)\\
\ell_B^{(n)}&\geq  \max\left\{0,{I^{{\eps}_{B}}_{\infty}[U^n;Y^n]}+1\right\}+ \log\ln(1/\eps_B).
\end{align*}
It thus follows from Theorem \ref{jaikumar's proof} that for all large $n$, there
exists a pair of encoding functions $e_A^{(n)}$ and $e_B^{(n)}$ and a
decoding function $d_C^{(n)}$ such that
$\Pr\left\{X^n \neq d^{(n)}_C\left(e^{(n)}_A
(X^n),e^{(n)}_B(Y^n)\right)\right\} \leq \eps$. Thus, from Definition
\ref{sidddef} it now follows that the rate pair $(R_1,R_2)$ is
$\eps$-achievable. Note that we can claim similar results using Theorem \ref{oneshotsideinf}.

Next, to show the converse (that is, the necessity of conditions
(\ref{ineq:nech1})--(\ref{ineq:nech2}), assume that there exist encoding
functions $e^{(n)}_A$ and $e^{(n)}_B$ and decoding functions
$d^{(n)}_C$ such that for all large $n$, we have $nR_1\geq
\ell_A^{(n)}$, $nR_2 \geq \ell_B^{(n)}$ and
$\Pr\left\{X^n \neq d^{(n)}_C\left(e^{(n)}_A
(X^n),e^{(n)}_B(Y^n)\right)\right\} \leq \eps + \delta$.
From Theorem \ref{conversesideinf} we conclude that for all such $n$:
\begin{align*}
nR_1 &\geq \ell^{(n)}_A \geq {H_{0}^{\eps+\delta}[X^n|U^n]}\\
nR_2 &\geq \ell^{(n)}_B \geq I^{\eps+\delta}_{\infty}[U^n;Y^n] + \log\left(\eps+\delta\right).
\end{align*}
Our claim follows by dividing these equations by $n$ and taking ${\limsup}$ of both sides as $n\rightarrow \infty$.
\end{proof}
 
We will now show that Theorem \ref{mksw} implies the result of Miyake and
Kanya~\cite[Theorem 2]{miyakaye-kanaya-1995}. Towards this we will use Lemma \ref{assymcond134} and the following convergence result.

\begin{lemma} (Datta \cite{Datta-it-2009})
\label{asympinfinity}
 Let $\bP = \{P_n\}_{n=1}^{\infty}$ and $\bQ = \{Q_n\}_{n=1}^{\infty}$ be an arbitrary sequences of probability mass functions defined on the set $\{\cX^n\}_{n=1}^{\infty}$, where $\cX^n$ is the $n\mhyphen$th cartesian product of the set $\cX$ and $|\cX| < \infty$. Assume that for every $n \in \mathbb{N}$, $\Supp(P_n) \subseteq \Supp(Q_n)$. Then
\beq
\label{noniid}
\lim_{\eps \to 0} \limsup_{n \to \infty} \frac{1}{n}{D}^{\eps}_{\infty}(P_n||Q_n) = \overline{I}(\bP;\bQ), \nonumber
\enq
In particular, if $\bP = \{P^{\times n}\}_{n =1}^{\infty}$ and $\bQ = \{Q^{\times n}\}_{n =1}^{\infty}$, where $P^{\times n}$ and $Q^{\times n}$ represent the product distributions of $P$ and $Q$ on $\cX^n$. Then
\beq
\label{iid}
\lim_{\eps \to 0} \limsup_{n \to \infty} \frac{1}{n}{D}^{\eps}_{\infty}(P_n||Q_n)=D(P||Q).
\enq
\end{lemma}
The proof for the above lemma was given by Datta in \cite{Datta-it-2009} for the quantum case. We give the proof in the classical case for the sake of completeness in the appendix. 
\begin{theorem}
The rate pair $(R_1, R_2)$ is achievable if and only if
\begin{align*}
R_1 &\geq \oH[\bX|\bU]\\
R_2 &\geq\overline{I}[\bU;\bY],
\end{align*}
where the sequence of random variables $U^n$ satisfies the Markov chain condition $p_{X^nY^nU^n}=p_{X^nY^n}p_{U^n\mid Y^n}$ for each $n$ (
Note that this is just a restatement of the result of Miyake and
Kanya~\cite[Theorem 2]{miyakaye-kanaya-1995}).

\end{theorem}
\begin{proof}
To establish the ``if'' direction, it is enough to
show that for all $(R_1, R_2)$ such that
\begin{align*}
R_1 &> \oH[\bX|\bU]\\
R_2 &> \overline{I}[\bU;\bY].
\end{align*}
and all $\eps >0$, $(R_1,R_2)$ is $\eps$-achievable for all large $n$.
Fix $\eps > 0$. Since $H_{0}^{\eps_A}$ is monotone non-increasing in
$\eps_A$ and similarly $I^{\eps_B}_\infty$ is a monotone non-increasing in $\eps_B$, where $2\eps_A + 4\eps_B \leq \eps$, we conclude from Lemma \ref{assymcond134} and Lemma~\ref{asympinfinity} that for all large
$n$,  inequalities (\ref{ineq:suffh1})--(\ref{ineq:suffh2}) are satisfied.
Our claim follows immediately from part (a). 

For the ``only if'' direction fix a rate pair $(R_1,R_2)$ and a pair
of encoding functions $e^{(n)}_{A}$ and $e^{(n)}_{B},$ where
$e^{(n)}_A: \cX^n \to [1: 2^{\ell_A^{(n)}}]$ and $e^{(n)}_B: \cY^n \to
[1: 2^{\ell^{(n)}_B}]$, and a decoding function $d^{(n)}_C : [1:
  2^{\ell_A^{(n)}}]\times [1: 2^{\ell_B^{(n)}}] \to \cX^n \times
\cY^n$, satisfying Definition~\ref{sidddef} with $\eps=0$. Fix
$\gamma>0$, and using the necessary condition in part (a), with
$\eps\leftarrow \frac{\gamma}{2}$ and $\delta \leftarrow
\frac{\gamma}{2}$, conclude that
\begin{align}
R_1&\geq  \limsup_{n \to \infty} \frac{H_{0}^{\gamma}[X^n|U^n]}{n}\\
R_2&\geq  \limsup_{n \to \infty} \frac{I_{\infty}^{\gamma}[U^n;Y^n]}{n}
\end{align}
Our claim follows by taking limits as
$\gamma$ goes to $0$, and using Lemma \ref{assymcond134} and Lemma~\ref{asympinfinity}.
\end{proof}

\section{One-shot source coding under maximum distortion criterion}
\subsection{Proof of Theorem \ref{distac}}

\begin{proof}
Let us first assume that $I^{\eps_1}_{\infty}[X;Y] >0$. We will deal with the case when $I^{\eps_1}_{\infty}[X;Y]<0$ separately.\\

\noindent {\bf{Random codebook generation:}}
 Generate $Y_1,\dots,Y_{2^{\ell_A}} \in \cY$ independently with probability distribution $p_Y$, where $p_Y$ is the marginal of $p_{XY}$. These $2^{\ell_A}$ independent realizations of $Y$ forms a random codebook, i.e., 
\beq
\cC = \{Y_1,\dots,Y_{2^{\ell_A}}\}. \nonumber
\enq 
{{\bf{Encoding:}}
For every $x \in \cX$ we define the encoder $e$ by $e(x)= i$, where $i$ is determined from
\beq
d(x,Y_i) = \min_{1\leq j \leq 2^{\ell_A}} d(x, Y_j). \nonumber  
\enq
}
{{\bf{Decoding:}}
If the decoder receives $i$, then the decoder $f$ outputs $Y_i$, i.e., 
\beq
f(i) = Y_i.  \nonumber
\enq
}
{\bf{Analysis for the distortion:}} We will now show that for our encoding and decoding and choice of $\ell_A$ (as mentioned in the statement of the theorem) we have $\Pr\{d(X,f(e(X))) > \bar{\lambda}_{\eps_{1}}(X,Y)\} \leq \eps$. This would further imply that  
\beq
\bar{\lambda}_{\eps}(X,f(e(X)))\leq\bar{\lambda}_{\eps_1}(X,Y). \nonumber
\enq
Let us now define
\begin{align}
\label{distset}
\Gamma^{(\eps_1)} &:= \{(x,y):d(x,y) \leq \bar{\lambda}_{\eps_1}(X,Y)\}.
\end{align}
Let $\phi \in \cB^{\eps_1}(P)$ be such that
\begin{align}
\label{optphi2} 
I^{\eps_1}_{\infty}[X;Y] = \log \max_{(x,y) \in \cX \times \cY} \frac{\phi(x,y)}{p_{X}(x)P_{Y}(y)}.
\end{align}
Further, let us define the following indicator function which will help us in the calculations below
\beq
\mbox{\large$\chi$}(x,y) =
\begin{cases}
1 & \mbox{if } (x,y) \in \Gamma^{(\eps_1)},\\
0        & \mbox{otherwise}.
\end{cases}
\enq
\begin{align*}
\Pr\left\{d(X,f(e(X))) >\bar{\lambda}_{\eps_1}(X,Y)\right\}&=\Pr\left\{d(X,Y_i) >\bar{\lambda}_{\eps_1}(X,Y) , \forall i \in [1:2^{\ell_A}] \right\}\\
&=\sum_{x\in\cX}p_{X}(x)\Pr\bigg\{d(x,Y_i)>\bar{\lambda}_{\eps_1}(X,Y)~ \forall i \in [1:2^{\ell_A}] \bigg\}\\
&\overset{a} = \sum_{x\in\cX}p_{X}(x)\bigg(\Pr\bigg\{d(x,Y)>\bar{\lambda}_{\eps_1}(X,Y)\bigg\}\bigg)^{2^{\ell_A}}\\
&\leq \sum_{x\in\cX}p_{X}(x)\left(1-\sum_{y\in\cY}P_{Y}(y)\mbox{\large$\chi$}(x,y)\right)^{2^{\ell_A}}\\
&\overset{b}\leq \sum_{x\in\cX}p_{X}(x) \bigg(1-2^{-I^{\eps_1}_{\infty}[X;Y]}\sum_{{y}\in \cY}\frac{\phi(x,y)}{p_X(x)}\mbox{\large$\chi$}(x,y)\bigg)^{2^{\ell_A}}\\
&\overset{c}\leq  \sum_{x\in\cX}p_{X}(x)e^{-\bigg(2^{\ell_A}2^{-I^{\eps_1}_{\infty}[X;Y]}\sum_{{y}\in \cY}\frac{\phi(x,y)}{p_X(x)}\mbox{\large$\chi$}(x,y)\bigg)}\\
&\overset{d} \leq \sum_{x\in\cX}p_{X}(x) \left(1 - \sum_{{y}\in \cY}\frac{\phi(x,y)}{p_X(x)}\mbox{\large$\chi$}(x,y)\right)+ e^{-2^{\ell_A}2^{-I^{\eps_1}_{\infty}[X;Y]}}\\
& = 1- \sum_{(x,y) \in \Gamma^{(\eps_1)}}\phi(x,y) + e^{-2^{\ell_A}2^{-I^{\eps_1}_{\infty}[X;Y]}}\\
& \overset{e}\leq 2\eps_1 + e^{-2^{\ell_A}2^{-I^{\eps_1}_{\infty}[X;Y]}},
\end{align*}
where $a$ follows because $Y_1,\dots,Y_{2^{\ell_A}}$ are independent and identically distributed according to $P_Y$;  $b$ follows because for every $(x,y) \in \mathcal{X} \times\cY$ we have $p_{Y}(y)\geq 2^{-I^{\eps_1}_{\infty}[X;Y]}\frac{\phi(x,y)}{p_X(x)}$; $c$ follows from the inequality $(1-x)^y \leq e^{-xy} ~ (0 \leq x \leq 1, y\geq 0)$ in our setting $x = 2^{-I^{\eps_1}_{\infty}[X;Y]}\sum_{y:(x,y) \in \Gamma^{(\eps_1)}}\frac{\phi(x,y)}{p_X(x)}$ and $y =2^{\ell_A}$; $d$ follows because of the inequality $e^{-xy} \leq 1-x+e^{-y} ~(0\leq x \leq 1, y \geq 0)$ in our setting $x = \sum_{y:(x,y) \in \Gamma^{(\eps_1)}} \frac{\phi(x,y)}{p_X(x)}$ and $y =2^{\ell_A}2^{-I^{\eps_1}_{\infty}[X;Y]}$ and $e$ follows because of the following set of inequalities 
 \begin{align}
1-\eps_1 & \overset {a}\leq \sum_{(x,y) \in \cX \times \cY} \phi(x,y) \nonumber\\
 & = \sum_{(x,y) \in \Gamma^{c(\eps_1)}} \phi(x,y)+  \sum_{(x,y) \in \Gamma^{(\eps_1)} }\phi(x,y) \nonumber\\
 & \overset{b} \leq \Pr\{\Gamma^{c(\eps_1)}\} +\sum_{(x,y) \in \Gamma^{(\eps_1)} }\phi(x,y) \nonumber\\
 \label{phibound}
 1- \eps_1&\overset{c} \leq \eps_1 + \sum_{(x,y) \in \Gamma^{(\eps_1)} }\phi(x,y),
 \end{align}
 where $a$ and $b$ both follow from the fact that, $\phi(x,y) \in \cB^{\eps_1}(P_{XY})$ and $c$ follows \eqref{distset}. Thus, we can now conclude that $\bar{\lambda}_{\eps}(X,f(e(X))) \leq \gamma$ if
\beq
{\ell_A} \geq I^{\eps_1}_{\infty}[X;Y] +\log[-\ln (\eps-2\eps_1)]. \nonumber 
\enq 

We now consider the case when $D^{\eps_1}_{\infty}(P_{XY}||P_X\times P_Y)< 0$. We will show that ${\ell_A} = 0$ is sufficient enough to satisfy the $\eps$-maximum distortion bound. An equivalent way of proving this claim is to prove the following
\beq
\label{k0}
\Pr\bigg\{(X, Y_1) \notin \Gamma^{(\eps_1)}\bigg\} \leq \eps,
\enq
where $\Gamma^{(\eps_1)}$ is defined in \eqref{distset} and $(X,Y_1) \sim P_XP_Y$.
To prove \eqref{k0} notice that
 \beq
 \label{neg}
D^{\eps_1}_{\infty}(P_{XY}||P_X\times P_Y) < 0.
 \enq
further implies
\beq
\label{ratiobound}
\frac{\phi(x,y)}{p_{X}(x)P_{Y}(y)} \leq 1, ~ \forall (x,y) \in \cX \times \cY.
\enq 
Let us now calculate the following
\begin{align}
\Pr\bigg\{(X, Y_1) \in \Gamma^{(\eps_1)}\bigg\} &\overset{a}= \sum_{(x,y) \in \Gamma^{(\eps_1)}} p_{X}(x)P_{Y}(y) \nonumber\\
&\overset{b} \geq \sum_{(x,y) \in \Gamma^{(\eps_1)}} \phi(x,y) \nonumber\\
& \overset{c} \geq 1-2\eps_1 \nonumber\\ 
\label{last11}
&\overset{d} > 1-\eps,
\end{align}
where $a$ follows because $Y_1 \sim P_Y$ is generated independent of $X$; $b$ follows from \eqref{ratiobound}; $c$ follows from \eqref{phibound} and $d$ follows from the fact that $2\eps_1<\eps$. \eqref{k0} now trivially follows from \eqref{last11}. From \eqref{k0} it now easily follows that ${\ell_A}=0$, would suffice for the bound
\beq
\bar{\lambda}_{\eps}(X,f(e(X)))\leq \gamma, \nonumber
\enq
to hold.
\end{proof}
\subsection{Proof of Theorem \ref{distconv}}
Let us define the following set
\beq
\label{settrick1}
\cA := \left\{y \in \cY: P_{Y}(y) < \frac{\eps}{2^{\ell_A}}\right\},
\enq
where $\cY := [1:2^{\ell_A}]$. Notice that $Y = f(e(X))$ cannot take more than $|e|$ values where $|e|$ denotes the size of the range of the encoder $e$. It is now easy to see from our construction of the set $\cA$ that $\Pr\{\cA\} < \eps$. Let $\phi(\cdot,\cdot)$ be a positive function such that
\beq
\label{phiconverse1}
\phi(x,y) :=
\begin{cases}
P_{XY}(x,y) & \mbox{if}~ y \notin \cA ,\\ 
0        & \mbox{otherwise}.
\end{cases}
\enq
Thus from \eqref{phiconverse1} and the fact that $\Pr\{\cA\} < \eps$, we have $\sum_{(x,y)\in \cX\times\cY}\phi(x,y) \geq 1-\eps$. Let $(x^*,y^*) \in \cX\times \cY$ be such that
\beq
\label{optiphi1}
\frac{\phi(x,y)}{p_{X}(x)P_{Y}(y)} \leq \frac{\phi(x^*,y^*)}{p_{X}(x^*)P_{Y}(y^*)}~~ \forall (x,y) \in \cX \times \cY.
\enq

We now have the following set of inequlities
\begin{align*}
I^{\eps}_{\infty}[X;Y]& \overset{a}\leq \max_{(x,y)\in \cX\times\cY}\log \frac{\phi(x,y)}{p_{X}(x)P_{Y}(y)}\\
&\overset{b}=\log\frac{\phi_{XY}(x^*,y^*)}{p_{X}(x^*)P_{Y}(y^*)}\\
& \overset{c} = \log\frac{P_{XY}(x^*,y^*)}{p_{X}(x^*)P_{Y}(y^*)}\\
&= \log\frac{P_{Y|X}(y^*|x^*)}{P_{Y}(y^*)}\\
& \overset{d} \leq \log\frac{1}{P_{Y}(y^*)}\\
& \overset{e} \leq \log\frac{2^{\ell_A}}{\eps}\\
& = -\log\eps+  \ell_A,
\end{align*}
 where $a$ follows from Definition \ref{smoothorderinf}; $b$ follows from \eqref{optiphi1}; $c$ follows from \eqref{phiconverse1}; $d$ follows because $P_{Y|X}(y^*|x^*)\leq 1$ and $e$ follows from \eqref{settrick1}. This completes the proof. 
\section{Conclusion and Acknowledgement}
In conclusion we have shown that many information theoretic results can be proven without invoking lemmas which are based on AEP. We have further shown that the smooth \renyi quantities play an intimate role in exhibiting achievability and converse proofs in the one-shot regime. Thus, giving an operational meaning to these quantities. Furthermore, we show that the bounds obtained using these smooth \renyi quantities converge to the results that are known in the asymptotic setting in the limit of large blocklength and when the error goes to zero. 

The author gratefully acknowledges the helpful discussions with Prof. Jaikumar Radhakrishnan.

\section*{Appendix}
 \subsection*{\underline{Proof of Lemma \ref{asympinfinity}}}
 We will first prove 
\beq
\lim_{\eps \to 0} \limsup_{n \to \infty} \frac{1}{n}D^{\eps}_{\infty}(P_n||Q_n) \leq \Ib(\bP;\bQ). \nonumber
\enq
Consider any $\lambda > \Ib(\bP;\bQ)$. Let us define the following set 
\beq
\label{consset}
\cA_{n}(\lambda) := \left\{\bx : \frac{1}{n} \log \frac{P_n(\bx)}{Q_n(\bx)} \leq \lambda\right\}. 
\enq
Let $\phi_{n}: \cX^n \to [0,1]$, $n \in \mathbb{N}$, such that 
\beq
\label{constsupinfrate}
\phi_n(\bx) =
\begin{cases}
P_{n}(\bx) & \mbox{if } \bx \in \cA_{n}(\lambda), \\
 0      & \mbox{otherwise.}
\end{cases}
\enq
From  Definition \ref{asyminftydiv} it easily follows that 
\beq 
\lim_{n \to \infty} \Pr\{\cA_{n}(\lambda)\} =1.
\enq
Thus from our  construction of $\phi_n$, \eqref{constsupinfrate}, it follows that
\beq
\label{constsminf}
\lim_{n \to \infty}\sum_{\bx \in \cX^n} \phi_n(\bx) = \lim_{n \to \infty} \Pr\{\cA_{n}(\lambda)\} =1.
\enq 
Notice the following inequalities
\begin{align*}
\lim_{\eps \to 0} \limsup_{n \to \infty}\frac{1}{n}D^{\eps}_{\infty}(P_n||Q_n) &\overset{a} \leq \limsup_{n\to \infty}D_{\infty}(\phi_n||Q_n)\\
& = \limsup_{n \to \infty} \frac{1}{n} \log \max_{\bx \in \cX^n} \frac{\phi_n(\bx)}{Q_{n}(\bx)}\\
&\overset{b} \leq \lambda,
\end{align*}
where $a$ follows from Definition \ref{smoothorderinf} and \eqref{constsminf}; $b$ follows from \eqref{consset} and \eqref{constsupinfrate}.

We now prove the other direction, i.e., 
\beq
\lim_{\eps \to 0} \limsup_{n \to \infty} \frac{1}{n}D^{\eps}_{\infty}(P_n||Q_n) \geq \Ib(\bP;\bQ). \nonumber
\enq
For every $\eps >0$, consider a sequence of positive functions $\{\phi_{n}\}_{n =1}^\infty$, such that for every $n \in \mathbb{N}$, $\phi_n \in \cB^{\eps}(P_n)$ and $D^{\eps}_{\infty}(P_n||Q_n)=\max_{\bx \in\cX^n}\log\frac{\phi_n(\bx)}{Q_n(\bx)}$.

Let $f(\eps)= \limsup_{n\to\infty}\frac{1}{n}D^{\eps}_{\infty}(P_n||Q_n)$. Then, from the definition of limit superior it follows that for every $\delta >0$ there exists $n_0$ such that for every $n >n_0$ we have $\frac{1}{n}D^{\eps}_{\infty}(P_n||Q_n)< f(\eps)+\delta$. Let 
\beq
\label{conssetasym0}
\cD_{n} := \left\{\bx: \frac{1}{n} \log\frac{P_n(\bx)}{Q_{n}(\bx)} > f(\eps)+\delta \right\}.
\enq
We will now show that for $n$ large enough and for every $\beta >0,$
\beq
\sum_{\bx \in \cD_{n}}P_{n}(\bx) \leq \eps+\beta. 
\enq
Towards this notice the following set of inequlities
\begin{align}
&\sum_{\bx \in \cD_{n}}P_{n}(\bx) \nonumber\\
&= \sum_{\bx \in \cD_{n}}\left[P_{n}(\bx)-\phi_{n}(\bx)\right] + \sum_{\bx \in \cD_{n}}\phi_n(\bx) \nonumber\\
&= \sum_{\bx \in \cD_{n}}\left[P_{n}(\bx)-\phi_{n}(\bx)\right]+ 2^{n\left(f(\eps)+\frac{\delta}{2}\right)}\sum_{\bx \in \cD_{n}}Q_{n}(\bx)
\nonumber\\
&\hspace{5mm}+\sum_{\bx \in \cD_{n}}\left[\phi_{n}(\bx)-2^{n\left(f(\eps)+\frac{\delta}{2}\right)}Q_{n}(\bx)\right] \nonumber\\
&\overset{a}\leq \eps+2^{{n\left(f(\eps)+\frac{\delta}{2}\right)}-n\left(f(\eps)+\delta\right)}\nonumber\\
&\hspace{5mm}+ \sum_{\bx:\phi_{n}(\bx)\geq2^{n\left(f(\eps)+\frac{\delta}{2}\right)}Q_{n}(\bx)}\left[\phi_{n}(\bx)-2^{n\left(f(\eps)+\frac{\delta}{2}\right)}Q_{n}(\bx)\right] \nonumber\\
&\overset{b}= \eps+2^{-n\frac{\delta}{2}}\nonumber\\
\label{mainprob}
&\overset{c}\leq \eps+\beta
\end{align}
where $a$ follows from the fact that $\phi_{n} \in \cB^\eps(P_n)$ and from \eqref{conssetasym0}; $b$ follows from the definition of $D^{\eps}_{\infty}(P_n||Q_n)$ and the fact that for $n$ large enough $\frac{1}{n}D^{\eps}_{\infty}(P_n||Q_n)< f(\eps)+\frac{\delta}{2}$ and $c$ follows from the fact that for $n$ large enough we have $2^{-n\frac{\delta}{2}} \leq \beta$. Thus, from \eqref{mainprob} it now follows that for $n$ large enough for any $\beta >0$, and in the limit $\eps \to 0$, we have
\beq
\label{finprob1}
\sum_{\bx \in \cD_{n}}P_{n}(\bx) \leq \beta.
\enq
Thus, from \eqref{finprob1} and the definition of $\Ib(\bP;\bQ)$ it now follows that
\beq
\lim_{\eps \to 0} \limsup_{n \to \infty} \frac{1}{n}D^{\eps}_{\infty}(P_n||Q_n) \geq \Ib(\bP;\bQ). \nonumber
\enq
The claim in \eqref{iid} easily follows from the law of large numbers and Definition \ref{asyminftydiv}. This completes the proof

\bibliographystyle{ieeetr}
\bibliography{References}
\end{document}

%% file: Slepianwolf.pdf_t
\begin{picture}(0,0)%
\includegraphics{Slepianwolf.pdf}%
\end{picture}%
\setlength{\unitlength}{4144sp}%
\begingroup\makeatletter\ifx\SetFigFont\undefined%
\gdef\SetFigFont#1#2#3#4#5{%
  \reset@font\fontsize{#1}{#2pt}%
  \fontfamily{#3}\fontseries{#4}\fontshape{#5}%
  \selectfont}%
\fi\endgroup%
\begin{picture}(10785,4158)(-374,-6640)
\put(-359,-3571){\makebox(0,0)[lb]{\smash{{\SetFigFont{25}{30.0}{\rmdefault}{\mddefault}{\updefault}{\color[rgb]{.5,.17,0}$\bm{X}$}%
}}}}
\put(-359,-5821){\makebox(0,0)[lb]{\smash{{\SetFigFont{25}{30.0}{\rmdefault}{\mddefault}{\updefault}{\color[rgb]{.5,.17,0}$\bm{Y}$}%
}}}}
\put(10396,-4741){\makebox(0,0)[lb]{\smash{{\SetFigFont{25}{30.0}{\rmdefault}{\mddefault}{\updefault}{\color[rgb]{.5,.17,0}$\bm{\left(\hat{X},\hat{Y}\right)}$}%
}}}}
\put(3691,-3301){\rotatebox{26.5}{\makebox(0,0)[lb]{\smash{{\SetFigFont{25}{30.0}{\rmdefault}{\mddefault}{\updefault}{\color[rgb]{.5,.17,0}$\bm{M_A \in \left\{1,\cdots,2^{\ell_A}\right\}}$}%
}}}}}
\put(3601,-6001){\rotatebox{332.0}{\makebox(0,0)[lb]{\smash{{\SetFigFont{25}{30.0}{\rmdefault}{\mddefault}{\updefault}{\color[rgb]{.5,.17,0}$\bm{M_B \in \left\{1,\cdots, 2^{\ell_B}\right\}}$}%
}}}}}
\end{picture}%

%% file: helper.pdf_t
\begin{picture}(0,0)%
\includegraphics{helper.pdf}%
\end{picture}%
\setlength{\unitlength}{4144sp}%
\begingroup\makeatletter\ifx\SetFigFont\undefined%
\gdef\SetFigFont#1#2#3#4#5{%
  \reset@font\fontsize{#1}{#2pt}%
  \fontfamily{#3}\fontseries{#4}\fontshape{#5}%
  \selectfont}%
\fi\endgroup%
\begin{picture}(10785,4158)(-374,-6640)
\put(-359,-3571){\makebox(0,0)[lb]{\smash{{\SetFigFont{25}{30.0}{\rmdefault}{\mddefault}{\updefault}{\color[rgb]{.5,.17,0}$\bm{X}$}%
}}}}
\put(-359,-5821){\makebox(0,0)[lb]{\smash{{\SetFigFont{25}{30.0}{\rmdefault}{\mddefault}{\updefault}{\color[rgb]{.5,.17,0}$\bm{Y}$}%
}}}}
\put(3691,-3301){\rotatebox{26.5}{\makebox(0,0)[lb]{\smash{{\SetFigFont{25}{30.0}{\rmdefault}{\mddefault}{\updefault}{\color[rgb]{.5,.17,0}$\bm{M_A \in \left\{1,\cdots,2^{\ell_A}\right\}}$}%
}}}}}
\put(3601,-6001){\rotatebox{332.0}{\makebox(0,0)[lb]{\smash{{\SetFigFont{25}{30.0}{\rmdefault}{\mddefault}{\updefault}{\color[rgb]{.5,.17,0}$\bm{M_B \in \left\{1,\cdots, 2^{\ell_B}\right\}}$}%
}}}}}
\put(10396,-4741){\makebox(0,0)[lb]{\smash{{\SetFigFont{25}{30.0}{\rmdefault}{\mddefault}{\updefault}{\color[rgb]{.5,.17,0}$\bm{\hat{X}}$}%
}}}}
\end{picture}%

%% file: single.pdf_t
\begin{picture}(0,0)%
\includegraphics{single.pdf}%
\end{picture}%
\setlength{\unitlength}{4144sp}%
\begingroup\makeatletter\ifx\SetFigFont\undefined%
\gdef\SetFigFont#1#2#3#4#5{%
  \reset@font\fontsize{#1}{#2pt}%
  \fontfamily{#3}\fontseries{#4}\fontshape{#5}%
  \selectfont}%
\fi\endgroup%
\begin{picture}(14835,1953)(-5144,-5020)
\put(-5129,-4201){\makebox(0,0)[lb]{\smash{{\SetFigFont{25}{30.0}{\rmdefault}{\mddefault}{\updefault}{\color[rgb]{.5,.17,0}$\bm{X}$}%
}}}}
\put(811,-3841){\makebox(0,0)[lb]{\smash{{\SetFigFont{25}{30.0}{\rmdefault}{\mddefault}{\updefault}{\color[rgb]{.5,.17,0}$\bm{M_A\in \left\{1,\cdots,2^{\ell_A}\right\}}$}%
}}}}
\put(9676,-4246){\makebox(0,0)[lb]{\smash{{\SetFigFont{25}{30.0}{\rmdefault}{\mddefault}{\updefault}{\color[rgb]{.5,.17,0}$\bm{{Y}}$}%
}}}}
\end{picture}%